\documentclass[pra,twocolumn,nofootinbib]{revtex4}
\usepackage[bookmarks = true, pdfpagemode = None, pdfstartview = FitH, colorlinks = true, urlcolor = blue,hyperfootnotes=false]{hyperref}
\usepackage{hypcap}
\usepackage{mathrsfs}

\newcommand{\leftexp}[2]{{\vphantom{#2}}^{#1}\!{#2}}
\newcommand{\CX}{\leftexp{C}{X}}
\newcommand{\po}[1]{p_{\scriptscriptstyle\hspace{-.04em}{#1}}}
\newcommand{\pt}[2]{p_{\scriptscriptstyle\hspace{-.04em}{#1}\hspace{-.06em}{#2}}}

\newcommand{\loc}{\mathcal}

\usepackage{rotating}

\input{Qcircuit}
\xyoption{curve}

\begin{document}
\title{Fault-Tolerant Thresholds for Encoded Ancillae with Homogeneous Errors}
\author{Bryan Eastin}
\email{beastin@unm.edu}
\affiliation{Department of Physics and Astronomy, University of New Mexico, Albuquerque, NM 87131-1156}
\begin{abstract}
I describe a procedure for calculating thresholds for quantum computation
as a function of error model given the availability of ancillae prepared
in logical states with independent, identically distributed errors.  The
thresholds are determined via a simple counting argument performed on a
single qubit of an infinitely large CSS code.  I give concrete examples of
thresholds thus achievable for both Steane and Knill style fault-tolerant
implementations and investigate their relation to threshold estimates in
the literature.
\end{abstract}



\maketitle
\section{Introduction}
The threshold for quantum computation is defined as the error rate below
which, given a number of qubits that scales polynomially in the length of
the computation, the problem size, and the desired fidelity, it is
possible to implement an arbitrary quantum algorithm.  Knowledge of
thresholds is clearly an important tool for use in the design of quantum
computing architectures, but as of yet no simple, unified scheme exists
for determining them.  The great variance of threshold values in the
literature is due partly to advances and improvements in the field of
fault tolerance, i.e. error-conscious design, but also partly to differing
approximations and assumptions regarding error models and resources.  This
makes direct comparison of fault-tolerant procedures difficult and tends
to limit error models to some form of the depolarizing channel.  The work
presented herein was motivated by the desire for a method of threshold
estimation that could easily be applied to a variety of error models and
fault-tolerant methods.

One of the keys to achieving fault tolerance is the construction of
low-error encoded states as aids to computation.  These ancillary states,
henceforth termed ancillae, provide a baseline against which to check for
errors and a resource for performing gates that cannot otherwise be
implemented in a robust way.  For the sake of fault tolerance, ancillae
are typically prepared in a manner that reduces the frequency of
correlated errors at the expense of increased overhead in terms of qubits
and applied gates.  Accounting for the vast array of possible methods of
ancilla construction has proven a major impediment to the design of a
generic scheme for threshold estimation.

Recent work by Reichardt~\cite{Reichardt04} has indicated that, by
discarding states which show symptoms of error, moderately sized ancillae
can be prepared so well that their contribution to the failure probability
of an encoded circuit becomes small.  Moreover, my own investigations
suggest that the residual error on ancillae that have passed this sort of
inspection is primarily due to the verification step.  When the
verification circuit is fault tolerant, this implies that the residual
error in the inspection basis is uncorrelated.  Motivated by these
indications, I sought a way of determining the threshold given the
availability of ancillae with independent, identically distributed errors,
expecting that the result would not differ too much from threshold estimates
obtained using liberal qubit expenditure during ancilla production.

This paper describes such a means of threshold calculation for fault-tolerant methods employing CSS codes.  Much of the material presented here
can be regarded as an elaboration of work by Knill, particularly
Refs.~\cite{Knill04,Knill05}.  As in Ref.~\cite{Knill05}, I assume that it
is possible to prepare ancillae in logical (encoded) basis states such
that errors on the component qubits are independent and all qubits have
the same error spectrum.  Using this assumption as well as the structure
of CSS encoded gates and a few error propagation tricks, it is possible to
express the failure probability of an encoded operation in terms of the
error probabilities of a single strand of the code blocks.  In the limit
that the number of encoding qubits goes to infinity, the criterion for
encoded failure becomes particularly simple, consequently yielding a
threshold in which various error probabilities are free parameters,
including the probabilities of different kinds of Pauli errors on a
particular gate.

It should be stressed that the method described here does not constitute a
constructive procedure for fault-tolerant quantum computing.  Several details regarding syndrome decoding and the selection
of the quantum code are omitted, and, more importantly, no practical
prescription is given for preparing the appropriate ancillae. These
caveats do not invalidate the resulting thresholds, but, presently, the
difficulty of obtaining the requisite ancillae limits their
applicability.  The algorithm I describe is primarily useful either as a
metric for comparing the performance of fault-tolerant procedures under
diverse conditions or as a replacement for the numerical simulations
typically used to generate threshold estimates. This second application is
a particularly apt use for my threshold results because the simulations
become increasingly difficult as the size of the quantum code and the
proclivity to discard ancillae increase.  By contrast, the approximations
employed in deriving my algorithm increase in accuracy with the same
parameters.

The structure of this paper is as follows.  Section~\ref{sec:background}
introduces notation and gives a brief exposition of CSS codes, fault
tolerance, and thresholds.  Section~\ref{sec:assumptions} enumerates my
assumptions.  Section~\ref{sec:homogeneous} outlines the theory needed to
generate thresholds using an infinitely large CSS code and logical
ancillae with independent, identically distributed errors.
Section~\ref{sec:counting} describes how single line error probabilities
can be determined.  In \S\ref{sec:practicalities} details regarding the
implementation of my algorithm are considered along with the error models
to which it applies.  Section~\ref{sec:examples} performs this analysis
for a few cases of special interest, including Knill and Steane style
fault-tolerant quantum computation schemes paired with a selection of
error models.  Section~\ref{sec:finiteCodes} presents an alternative
algorithm for finite codes.  Finally, \S\ref{sec:conclusion}
concludes with a brief review of my results and a discussion of possible
directions for future work.

\section{Background\label{sec:background}}

\subsection{Notation}
Throughout this paper $X$, $Y$, and $Z$ are used to denote the Pauli
operators while $\CX$ denotes the controlled-NOT or \textsc{xor} gate.  I
also make use of Hadamard and phase gates, which are given in the standard
basis by
\begin{align}
H=\frac{1}{\sqrt{2}}
\left[
\begin{array}{cc}
1 & 1 \\
1 & -1
\end{array}
\right]
&&
\text{and}
&&
P=
\left[
\begin{array}{cc}
1 & 0 \\
0 & i
\end{array}
\right]
\end{align}
respectively.  Together these gates span the Clifford group, an important
subset of quantum operations.  The addition of the $\pi/4$ rotation,
\begin{equation}
T=
\left[
\begin{array}{cc}
1 & 0 \\
0 & e^{i\pi/4}
\end{array}
\right],
\end{equation}
completes a universal set for quantum computation.

I use quantum circuit notation rather extensively.  A review of this
formalism can be found in Nielsen and Chuang's book~\cite{NielsenChuang}.
I deviate slightly from their standard in that no decoration is added to
distinguish encoded circuits from unencoded circuits.

A group of qubits combined via a quantum code into a logical~(encoded)
state is referred to as a block.  Encoded states are identified by a line
over the contents of the ket; encoded gates are similarly decorated when
not part of a circuit diagram.

Operations which do not couple qubits that reside in the same block are
called transversal.  This designation is in contrast to some of the
literature, which insists additionally that transversal operations should
apply the same component operation to each qubit of a block.  In this
paper, operations satisfying both properties are called homogeneous.

\begin{figure}
\capstart
\xy \Qcircuit @R=.2em @C=.4em {
& & & & \ctrl{8} & \qw & \qw & \qw & \qw & \qw & \qw & \gate{H} & \meter \\
& & & & \qw & \ctrl{8} & \qw & \qw & \qw & \qw & \qw & \gate{H} & \meter \\
& & & & \qw & \qw & \ctrl{8} & \qw & \qw & \qw & \qw & \gate{H} & \meter \\
& \push{\ket{\bar{0}}\ } & & & \qw & \qw & \qw & \ctrl{8} & \qw & \qw & \qw & \gate{H} & \meter \\
& & & & \qw & \qw & \qw & \qw & \ctrl{8} & \qw & \qw & \gate{H} & \meter \\
& & & & \qw & \qw & \qw & \qw & \qw & \ctrl{8} & \qw & \gate{H} & \meter \\
& & & & \qw & \qw & \qw & \qw & \qw & \qw & \ctrl{8} & \gate{H} & \meter \\
\\
& & \qw & \qw & \targ & \qw & \qw & \qw & \qw & \qw & \qw & \qw & \qw & \qw & \qw & \qw & \qw \\
& & \qw & \qw & \qw & \targ & \qw & \qw & \qw & \qw & \qw & \qw & \qw & \qw & \qw & \qw & \qw \\
& & \qw & \qw & \qw & \qw & \targ & \qw & \qw & \qw & \qw & \qw & \qw & \qw & \qw & \qw & \qw \\
& & \qw & \qw & \qw & \qw & \qw & \targ & \qw & \qw & \qw & \qw & \qw & \qw & \qw & \qw & \qw \\
& & \qw & \qw & \qw & \qw & \qw & \qw & \targ & \qw & \qw & \qw & \qw & \qw & \qw & \qw & \qw \\
& & \qw & \qw & \qw & \qw & \qw & \qw & \qw & \targ & \qw & \qw & \qw & \qw & \qw & \qw & \qw \\
& & \qw & \qw & \qw & \qw & \qw & \qw & \qw & \qw & \targ & \qw & \qw & \qw & \qw & \qw & \qw
\gategroup{1}{3}{7}{3}{.1em}{\{}
}
\POS
,<21em,-4.7em>*\xycircle<5em,3em>{}="c"
;<15.3em,-6.4em>*{} ** @{-}
,"c";<17em,-14.5em>*{} ** @{-}
,<17.1em,-4em>
\Qcircuit @R=.5em @C=.4em {
& & & & \ctrl{1} & \gate{H} & \meter \\
& \qw & \qw & \qw & \targ & \qw & \qw & \qw & \qw & \qw
}
\endxy
\caption{A circuit showing the homogeneous application of several encoded
gates for the seven qubit code.  The inset on the right shows a single
strand which has been extracted from the coding blocks.  Note that, modulo
a label indicating the ancilla's starting state, the strand is identical
to the encoded circuit. \label{fig:transversal} }
\end{figure}
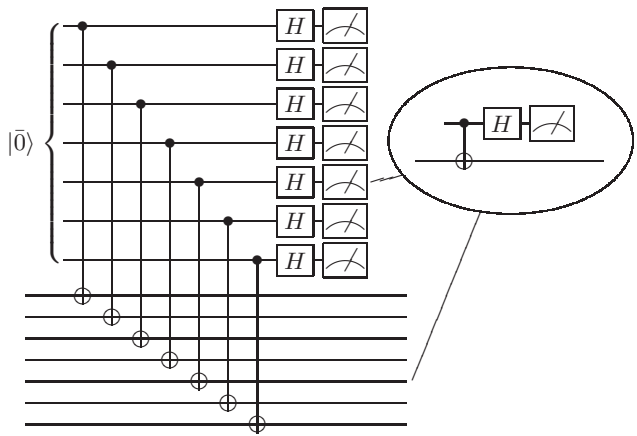

I refer to a single qubit of a block and all qubits that directly or
indirectly couple to it as a strand.  Gates between pairs of qubits should
be thought of as binding then into the same strand, as illustrated in
Fig.~\ref{fig:transversal}.

The notation $[[n,k,d]]$ represents a quantum code encoding $k$ logical
qubits using $n$ unencoded qubits and having a minimum distance of $d$.  A
code with minimum distance $d$ can correct any error of weight less than
or equal to $t=\lfloor\frac{d-1}{2}\rfloor$.  All codes dealt with in this
paper have $k=1$.

\subsection{CSS Codes}

A generalized Calderbank-Shor-Steane (CSS) code~\cite{PreskillNotes} is a
quantum code for which $X$ (bit flip) and $Z$ (sign flip) errors can be
corrected independently.  Canonical CSS codes have the additional property
that they are symmetric with respect to interchange of $X$ and $Z$.  This
symmetry implies that encoded $X$, $Y$, $Z$, $H$, and $\CX$ gates can be
implemented through transversal application of their unencoded
equivalents.  As a result of this useful property, CSS codes pervade the
literature on fault tolerance, including this paper.

\subsection{Fault Tolerance}
Fault-tolerant design is an approach to computation and error correction
that takes into account the fact that every component of a quantum
computer is likely to be unreliable.  Gates, ancillae, measurements, and
memory are all assumed to err with some probability.  Fault-tolerant
constructions seek to minimize the effect of these errors by preventing
their spread.  This is achieved through the use of transversal operations,
the expenditure of qubits, repetition, and teleportation.

The focus on minimizing the spread of errors is crucial.  To have any hope
of recovering corrupted data, whether it be classical or quantum, we need
some prior knowledge of the kinds of errors that are likely to occur.  In
the absence of any other information, it is generally assumed that errors
on different parts of a computer are uncorrelated.  Fault-tolerant
operations are designed so that the failure of a single component does not
result in an error on many different parts of an encoded state, thereby
avoiding the creation of correlated errors.

\subsection{Thresholds\label{subsec:thresholds}}
No matter how skillfully constructed the fault-tolerant procedure, there
remains for any finite code a nonzero probability that too many
independent errors will occur in a computational step and our data will
become irreparably corrupted.  Consequently, the probability of failure
approaches one as the length of the computation increases.  To ensure that
we can perform a computation of arbitrary length we need a way of making
the probability of an uncorrectable set of unencoded errors, i.e. an
encoded error, arbitrarily small.

The most common method of achieving an arbitrarily low encoded error rate
is known as concatenated coding.  In concatenated coding, the process of
encoding is divided into many levels.  Zeroth-level (physical) qubits are
used to encode first-level qubits, these first-level qubits are then in
turn used to encode second-level qubits, second-level qubits are used to
encode third-level qubits, and so on.  The basic idea is the following:
``If encoding qubits reduces the effective error rate then encoding the
encoded qubits should reduce the error rate even more.''  This provides us
with a plausible sounding way of achieving an arbitrarily small error
rate; we simply add layers of encoding until our error rate is acceptable.

But encoding will not always decrease our error rate.  It is possible for
our hardware to be so error prone that the process of applying an encoded
gate and error correcting is less likely to succeed than simply applying
the unencoded gate.  From this observation arises the idea of a threshold
error probability $p_{th}$ for quantum computation.  The threshold is the
unencoded error probability below which we can achieve an arbitrarily low
encoded error probability using a number of qubits that scales
polynomially in the size of the problem.  Put another way, it is the error
probability below which we can compute indefinitely.

Determining the threshold exactly for a given set of assumptions has
proven to be a hard problem, but we can get some idea of its value through
bounds and estimates.  Upper bounds resulting from proofs of classical
simulability have pushed the depolarizing threshold below
$50\%$~\cite{Buhrman06,Razborov03}, while lower bounds on the order of
$10^{-5}$~\cite{Aliferis06,Reichardt05} have been rigorously proven
through concatenation of explicit fault-tolerant constructions.  This
paper focuses on estimates of the threshold, which have recently settled
near the $1\%$ mark~\cite{Reichardt04,Knill04}.

Estimates of the threshold for quantum computation are generally made by
analyzing a particular fault-tolerant implementation using a specific
finite code under concatenation.  Fundamentally, these estimates derive
from the idea that encoding is undesirable if
\begin{equation}
\left\{
\parbox{5.3em}{\centering Encoded error rate}\rule{0em}{1.3em}
\right\}
>
\left\{
\parbox{5.5em}{\centering Unencoded error rate}\rule{0em}{1.3em}
\right\}.
\end{equation}
Intuitively, this makes sense; we would not expect error correction to be
advantageous when an encoded gate or qubit is more likely to fail than an
unencoded one.  Nonetheless, careful consideration of this justification
reveals some difficulties with the argument.  There are many sorts of
errors, and it might well be the case that the encoded error rate
increases for some of them, but not all.  Even were there only one kind of
error, however, the equation above would not provide us with a lower bound
on the threshold, but only an estimate.  While, in that case, an increase
in the error rate at the first level of concatenation implies that
subsequent layers of concatenation also increase the error rate, the
converse is not true.  To see why, assume we have some fault-tolerant
procedure for which the encoded failure rate is less than the unencoded
failure rate.  At the first level our code is constructed of unencoded
qubits that are either perfect or have failed.  At the second level of
encoding, however, our code is constructed of singly encoded qubits that
may be perfect, insufficiently corrupted to result in failure, or failed,
yet only the last case is considered an encoded error.  In some sense, the
qubits that have not failed are now of lower quality than they were at the
previous level.  Thus, the fact that encoding worked at the previous level
does not guarantee that it will work at the current one.

Details such as these imply that most threshold results are estimates rather than rigorous lower bounds.  The interested reader can find a more thorough discussion of this issue in work by Svore and others~\cite{Svore05,Svore06}.

\section{Assumptions\label{sec:assumptions}}
Some large part of the variance in fault-tolerant threshold calculations is
due to the variety of assumptions employed by various authors.  In an
effort to combat confusion, my assumptions are listed below in roughly
the order of decreasing novelty.
\begin{enumerate}
\item All ancillary qubits have independent, identical error distributions.\label{ass:ident}
\item There are no memory errors. \label{ass:nomemory}
\item $\CX$ is the only two-qubit gate. \label{ass:CXonly}
\item Any pair of qubits can interact via a two-qubit gate. \label{ass:nonlocal}
\item Error operators are trace preserving and lack systematic coherent terms. \label{ass:errtype}
\item Gate failures are uncorrelated. \label{ass:uncorrelated}
\item Classical computation is freely available. \label{ass:cheapclassical}
\end{enumerate}
Assumptions~\ref{ass:ident}, \ref{ass:nonlocal}, \ref{ass:errtype}, \ref{ass:uncorrelated}, and \ref{ass:cheapclassical} are necessary for my analysis; the others are convenient but optional.  Assumption~\ref{ass:nomemory} obviates the need to consider questions of parallelism, gate timing, and the speed of classical computation, while Assumption~\ref{ass:CXonly} reduces the number of cases that must be considered.

\section{Thresholds for Homogeneous Methods\label{sec:homogeneous}}
Two important observations from \S\ref{sec:background} provide the
foundation for my method of threshold calculation.  The first is that fault-tolerant procedures for CSS codes are, to a large degree, transversal.
The second is that, for the kind of CSS codes typically employed, these
transversal operations can be implemented by applying the same gate to
every qubit in a block.  The sum of these observations is that most
operations performed in a fault-tolerant procedure consist of doing the
same thing to each of the qubits in a block.  If one could arrange for all
operations to have this property, which I refer to henceforth as
homogeneity, analyzing the behavior of fault-tolerant circuits would be
greatly simplified.

Three components of the typical fault-tolerant method stand in the way of full
homogeneity: ancilla production, syndrome extraction, and recovery.  Starting from
the perspective of threshold estimation, this section addresses each of these aspects, partly by keeping in mind that the
eventual goal is to model errors, not computation.  Ultimately, the method derived is meaningful both as a form of threshold estimation and
as a threshold bound for idealized resources.

\subsection{Ancillae}
The ancillae used in CSS-code fault-tolerant procedures are typically
prepared in highly entangled states, i.e. in logical basis states.  By
definition, entangled states cannot be constructed without the interaction
of the constituent parts, so there is no a priori reason to think that the
qubits composing an ancilla will have either independent or identical
error distributions.  For fault-tolerant procedures, however, the
production of an entangled ancilla is usually followed by a homogeneous
verification circuit, and, in my experience~\cite{EastinResidualError},
most of the residual error probability (of the kind tested for) arises
during this verification step.  With this is mind, I approximate ancillae
as having uniform error distributions.

It is important to realize that this assumption is far less innocuous than
it sounds.  Implicitly, I am assuming that ancillae of the desired size
can be constructed for use in a verification circuit, but in subsequent
sections I take the limit $n\rightarrow\infty$.  Thresholds given in this
limit are only practically achievable if an efficient procedure exists to
prepare logical ancillae.  To be efficiently scalable, however, a
construction routine must have nonvanishing probability of generating an
ancilla that has good fidelity with the desired state.  Fault-tolerant
schemes using concatenated codes provide a method of achieving this for
arbitrarily large ancillae, but as a side effect of universal quantum
computation.  At present, there is no known method of preparing a logical
qubit encoded using a CSS code of arbitrary size that does not depend on
the ability to perform universal quantum computation.  Thus, absent an
explicit recipe for ancillae preparation, the algorithm presented in this
paper does not constitute a constructive procedure for achieving any
threshold.

\subsection{Error Location}
Current techniques for locating errors require performing a complicated
and distinctly non-homogeneous function on the output of ancilla
measurement.  But while this classical processing requires knowledge of
all the measurements, its effect, assuming that no more than the
correctable number of errors has occurred, can be described in terms
of the individual qubits.  So long as the total number of errors present
on a measured ancilla is less than half the minimum distance\footnote{Some
higher weight errors will also be correctable, but I only lower my
threshold by ignoring them.}, the effect of the classical
processing is to determine a subset of the measured bits that can be
flipped to yield an undamaged codeword.  For the purposes of
error correction, knowing this string is equivalent to knowing the
location of all the errors. While the second kind of information is not
directly available to a quantum computer, it is quite accessible to a
theorist treating errors probabilistically.  I can therefore model the
effect of classical processing in two steps.  First, I determine whether
too many errors have occurred on a block to permit proper decoding, and,
if this is not the case, I treat the location
of bit flips on the measured qubits as revealed.

I have reduced the process of error location to a non-homogeneous failure
check and an arguably homogeneous revelation step.  For a Monte-Carlo
simulation, the failure check would consist of polling all of the other
qubits and counting up the number of errors that have occurred to see
whether they exceeded half the minimum distance of the code.  If instead
we performed a probability flow analysis, the expected probability of
passing the check would be simply
\begin{equation}
E_\mathrm{p}(p_\loc{L}) = \sum_{i=0}^t\binom{n}{i}p_\loc{L}^i(1-p_\loc{L})^{n-i}\label{eq:ppass}
\end{equation}
where $n$ is the number of qubits in the block, $t$ is the maximum number
of errors that can be corrected with certainty by the code, and
$p_\loc{L}$ is the probability that a particular qubit has an $X$ error at
some location (step) $\loc{L}$, here chosen to be just after the time of
measurement.

Equation~(\ref{eq:ppass}) suggests a way of recovering full homogeneity.
Letting $\tau=t/n$, in the limit of large $n$, Eq.~(\ref{eq:ppass}) becomes
\begin{equation}
E_\mathrm{p}(p_\loc{L}) =
\begin{cases}
0 & \text{if $p_\loc{L}<\tau$} \\
1 & \text{if $p_\loc{L}>\tau$} \\
\end{cases}
\end{equation}
which is again homogeneous from the perspective of a simulation.  As an
added benefit, it is no longer necessary to concatenate many layers of
coding to achieve a rigorous threshold; instead a vanishing error
probability is achieved as the limit of a very large code.  This
alternative to concatenation is known as large block coding or, simply,
block coding.

The preceding paragraphs demonstrate that, for homogeneous~(independent,
identically distributed) errors, whether or not an encoded state on a
large number of qubits fails is determined by the error probability of an
individual qubit.  For this result to be useful, an infinite family of CSS
codes with nonvanishing fractional minimum distance must exist.
Fortunately, it has been shown~\cite{Steane96b,Calderbank96} that CSS
codes exist such that $\tau\approx 5.5\%$ for asymptotic values of $n$.
It is not known whether a similar claim can be made for CSS codes in which
the encoded phase gate can be implemented transversally, but this
convenience is not necessary for my construction.

The analysis of this section assumes minimum distance error correction,
but an identical result applies to error correction up to the channel
capacity. Gottesman and Preskill~\cite{Gottesman01} have shown that
families of general CSS codes exist that are asymptotically capable of
correcting errors up to $\tau\approx11\%$.  Hamada~\cite{Hamada04} has
shown that this result applies to CSS codes with $X$-$Z$ exchange symmetry
as well.

The appropriate choice for $\tau$ depends on the purpose of the calculation.
When the goal is to estimate the threshold that would be obtained by
running a Monte-Carlo simulation of a minimum distance decoder, $\tau$
should be chosen to be $5.5\%$.  To obtain the largest bound on the threshold
for homogeneous ancillae or for comparison to threshold estimates that use
the channel capacity, it is best to choose $\tau=11\%$. In other cases it
may be desirable to choose a value of $\tau$ specific to a family of
quantum error correcting codes with special properties, such as ease of
syndrome decoding or the possession of low weight stabilizer operators.

\subsection{Recovery}
Having diagnosed the location of our errors, the obvious way of dealing
with them is to apply to each qubit the gate which reverses its error.
Such a recovery operation is inherently inhomogeneous since not all qubits
will be in error, and thus not all qubits will have recovery gates applied
to them.  There are a number of ways to deal with this problem, but I
follow the lead of Knill~\cite{Knill05,Knill04} and dispense with recovery
altogether.  I can get away with this because a string of Pauli
errors can either be thought of as an error or as an operator shifting us
into a different~(but equally viable) code space.  Put another way, we can
ignore any errors that we know about since we know how to determine the
effect (see \S\ref{subsec:propagation}) they will have later in the
circuit and this effect is easily accounted for when analyzing the results
of measurements.

By a similar argument I need never apply any Pauli gates, including those
used to implement encoded Pauli gates!

\section{Error Counting\label{sec:counting}}
In the previous section we saw how to modify fault-tolerant procedures based on CSS
codes so that they are fully homogeneous.  The advantage of doing this is that
the error probabilities of qubits within an encoded block then become independent and identical.
Since each strand is functionally identical, it suffices to determine the
error spectrum for one of them; the probability of an encoded failure
at any point can be predicted from the error probabilities of an individual strand
of the transversal procedure.  As the number of encoding qubits becomes large, the fraction of
qubits with a particular error approaches the expectation for that error.
In the limit that $n\rightarrow\infty$, we can say for certain whether our
procedure fails on any given step since, in that limit, the
probability of an encoded failure becomes a step function.  Thus, the threshold
is completely determined by the probability of an encoded failure, and the
probability of an encoded failure is completely determined by the error
probability of a single strand of the blocks. Therefore, in order to calculate
the threshold I need only determine the error probability on a single
strand at every point in the fault-tolerant circuit.  This can be
accomplished through a combination of error propagation and exhaustive
bookkeeping which I describe in the following subsections.

\subsection{Pauli Error Propagation\label{subsec:propagation}}
Pauli error propagation relies on the fact that the Pauli group is
invariant under conjugation by Clifford gates.  This implies that any
string of Pauli gates followed by a Clifford gate is equivalent to the
same Clifford gate followed by some (possibly different) string of Pauli
gates.  Consequently, it is possible to shuffle Pauli errors to the end of
a Clifford circuit, thus yielding a perfect outcome modified by the
resultant Pauli operators (see Fig.~\ref{fig:eprop}).

Furthermore, as explained in appendix~\ref{app:pi4gate}, fault-tolerant
procedures based on CSS codes need never apply non-Clifford gates to the
data qubits.  Encoded gates outside of the Clifford group are performed by teleporting
the data onto a specially prepared ancillae.  Since only Clifford gates
are applied, error propagation can be used to determine the effect of a given Pauli error
at any later point in the procedure.

\begin{figure}
\centerline{
\Qcircuit @R=.5em @C=.8em {
& \gate{X} & \ctrl{1} & \qw & \push{\rule{1em}{0em}} & & \ctrl{1} & \gate{X} & \qw \\
& \qw & \targ & \qw & \raisebox{1.5em}{=} & & \targ & \gate{X} & \qw
}
}
\vspace{1em}
\centerline{
\Qcircuit @R=.5em @C=.8em {
& \gate{X} & \gate{H} & \gate{P} & \ctrl{1} & \qw & & & \gate{H} & \gate{P} & \ctrl{1} & \gate{Z} & \qw \\
& \gate{Y} & \ctrl{1} & \gate{X} & \targ & \qw & \push{\rule{1em}{0em}} & & \ctrl{1} & \gate{X} & \targ & \gate{X} & \qw\\
& \gate{Z} & \targ & \targ & \gate{P} & \qw & \raisebox{1.5em}{=} & & \targ & \targ & \gate{P} & \gate{X} & \qw \\
& \qw & \gate{Y} & \ctrl{-1} & \gate{H} & \qw & & & \gate{Y} & \ctrl{-1} & \gate{H} & \gate{X} & \qw
}
}
\caption{Examples of Pauli propagation.  Equalities are up to an overall phase. \label{fig:eprop}}
\end{figure}
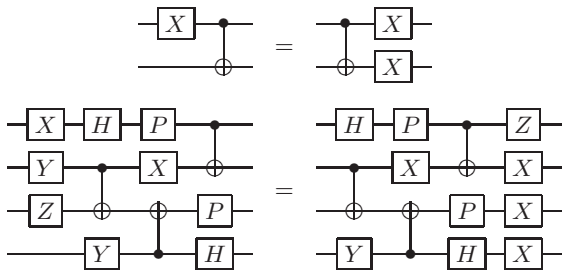

\subsection{Error Bookkeeping}
Given a gate, say the Hadamard, and a set of probabilities describing the
likelihood of various Pauli errors, say $\po{X}$, $\po{Y}$, and $\po{Z}$
for the errors $X$, $Y$, and $Z$, the post gate state can be written as a
probabilistically selected pure state, such that
\begin{equation}
\ket{\Psi^\prime}=
\begin{cases}
X H \ket{\Psi} & \text{with probability } \po{X}\\
Y H \ket{\Psi} & \text{with probability } \po{Y}\\
Z H \ket{\Psi} & \text{with probability } \po{Z}\\
H \ket{\Psi} & \text{otherwise.}
\end{cases}
\end{equation}
The effect of applying further Clifford gates is to change, via error propagation, which Pauli error corresponds to each probability, and then to add a second layer of probabilistic errors.  If, for example, we were to apply another Hadamard gate our state would become
\begin{equation}
\ket{\Psi^{\prime\prime}}=
\begin{cases}
X \ket{\Psi} & \text{with probability } \po{Z}+\po{X}+\po{Y} \po{Z}+\po{X}^2\\
Y \ket{\Psi} & \text{with probability } 2 \po{Y}+2 \po{X} \po{Z}\\
Z \ket{\Psi} & \text{with probability } \po{X}+\po{Z}+\po{Y} \po{X}+ \po{Z}^2\\
\ket{\Psi} & \text{otherwise.}
\end{cases}
\end{equation}
By repeated application of this process it is possible to determine the
probability of various kinds of errors at any point in a circuit composed
of Clifford gates.  Armed with this knowledge we can
determine\footnote{For non-transversal circuits the number of terms in our
bookkeeping rapidly becomes unmanageable as the size of the code
increases.} the likelihood of an encoded failure or, in the infinite
limit, whether an encoded failure will happen or not.

\section{Practicalities\label{sec:practicalities}}
While the previous section presented the basic algorithm for determining
whether an encoded failure occurs, this section deals with details of the error model
and the implementation that must be considered in any actual application
of the method.  Section~\ref{subsec:moreGenErr} identifies the error
models which can be adapted approximately to my method, a question which
has been considered for fault tolerance in general by several
authors~\cite{Aharonov99,Knill98,Preskill97}. Its conclusion, drawn by
Preskill in reference~\cite{Preskill97}, is that, roughly speaking, coherent errors with
random phases add like stochastic errors.  The finer points of how I
implement the algorithm are covered in \S\ref{sec:implementation}.

\subsection{More General Errors\label{subsec:moreGenErr}}
Stochastic Pauli errors are far from the only kind of error that can
affect a system, but they are an acceptable substitute for a variety of
other error channels.  To see why, begin by considering an arbitrary trace-preserving error operator $\mathcal{E}$.
The action, on a state $\rho$, of any such error operator can be written as
\begin{align}
\mathcal{E}(\rho) = \sum_j E_j\rho E_j^\dagger & & \mathrm{where} & & \sum_j E_j^\dagger E_j = I. \label{eq:QOpTrPr}
\end{align}
By interspersing errors of this form with perfect quantum gates it is possible to model any faulty quantum circuit that does not suffer from leakage.  When the error operators are local, it makes sense to approximate them by stochastic Pauli channels.  Given a local error operator satisfying Eq.~(\ref{eq:QOpTrPr}), I define the associated stochastic Pauli channel to have error probabilities
\begin{align}
\begin{split}
\po{X} &= \sum_j \left\lvert\mathrm{tr}(E_jX)\right\rvert^2, \\
\po{Y} &= \sum_j \left\lvert\mathrm{tr}(E_jY)\right\rvert^2\textrm{, and} \\
\po{Z} &= \sum_j \left\lvert\mathrm{tr}(E_jZ)\right\rvert^2.
\end{split}\label{eq:apstoch}
\end{align}
By design, this channel correctly reproduces the probability of measuring that a given Pauli error occurred after a single application of the general error operator.  Its suitability in more varied circumstances is the subject of the remainder of this subsection.

In practice, many gates are required (and therefore many error operators
act) between each error correction, so it is important to know how
errors accumulate.  As for the case of stochastic errors, general
trace-preserving errors can be separated from the associated circuit
providing that it is composed exclusively of Clifford gates.  The
separation is accomplished by applying Pauli propagation to each term in
the Pauli-basis decomposition of the elements, e.g. $E_j$, of the error
operator.  If the Clifford circuit is fault tolerant, then error
propagation maps single-qubit errors to single-qubit errors (on a given
encoded block).  Each of the resultant error operators differs from the actual error
by a simple relabeling of the local Pauli basis.  Thus, the circuit can be
disregarded; it is sufficient to consider how the transformed local errors accumulate.

A sequence of $s$ single-qubit trace-preserving errors acting on a state
$\rho$ can be written as
\begin{align}
\begin{split}
\mathcal{E}_s\circ&\cdots\circ\mathcal{E}_2\circ\mathcal{E}_1(\rho) \\
&= \sum_{j_s}\cdots\sum_{j_2}\sum_{j_1} \left(\prod_{k=s}^{1} E_{j_k k} \right)\rho\left(\prod_{k=1}^{s} E_{j_k k}^\dagger \right). \\
\end{split}
\end{align}
The probability of measuring, for example, an $X$ error on the resulting state is
\begin{align}
\po{\mathrm{E}X} &= \sum_{j_s}\cdots\sum_{j_2}\sum_{j_1} \left\lvert\mathrm{tr}\left(\prod_{k=s}^{1} E_{j_k k}X\right)\right\rvert^2. \label{eq:actXerr}
\end{align}
By contrast, replacing the error operators with their associated stochastic Pauli errors, as defined in Eq.~(\ref{eq:apstoch}), yields
\begin{align}
\po{\mathrm{S}X} &= \sum_{k=1}^s \sum_{j_k} \left\lvert\mathrm{tr}(E_{j_kk}X)\right\rvert^2 + O(p^2). \label{eq:appXerr}
\end{align}
When $\po{\mathrm{E}X}$ and $\po{\mathrm{S}X}$~(and the equivalent probabilities for $Y$ and $Z$) agree, the associated stochastic Pauli channel is a good substitute for the actual error channel.  To the lowest nontrivial order in $p$, the condition for equality can be derived as follows.

Consider each error operator $\mathcal{E}$ as a function of the total
single application error probability $p$.  Taylor expanding the
elements of $\mathcal{E}$ in $\sqrt{p}$ yields
\begin{align}
\begin{split}
E_0 &= I + \sqrt{p}(\alpha_0 X + \beta_0 Y + \gamma_0 Z + \delta_0 I) + O(p) \\
E_{j\neq0} &= \sqrt{p}(\alpha_j X + \beta_j Y + \gamma_j Z + \delta_j I) + O(p)
\end{split}\label{eq:TaylorErrOp}
\end{align}
where the freedom in the $E_j$ has been used to assure that $E_0$ contains
the only term independent of $p$.

Inserting expanded error operators of the form given in
Eq.~(\ref{eq:TaylorErrOp}) into Eq.~(\ref{eq:actXerr}) and discarding
terms of order greater than $p$ yields
\begin{align}
\begin{split}
\po{\mathrm{E}X} &\approx \sum_{k=1}^s\sum_{j_k\neq0} \left\lvert\mathrm{tr}\left(E_{j_k k}X\right)\right\rvert^2 + \left\lvert\mathrm{tr}\left(\prod_{k=s}^{1} E_{0 k}X\right)\right\rvert^2 \\
&\approx \sum_{k=1}^s\sum_{j_k\neq0} \left\lvert\mathrm{tr}\left(E_{j_k k}X\right)\right\rvert^2 + \left\lvert\sum_{k=1}^s\mathrm{tr}\left(E_{0 k}X\right)\right\rvert^2.
\end{split}\label{eq:appExXerr}
\end{align}
Thus, to first order in $p$, the difference between $\po{\mathrm{S}X}$ and
$\po{\mathrm{E}X}$ is
\begin{align}
\begin{split}
\po{\mathrm{E}X}-\po{\mathrm{S}X} &\approx \left\lvert\sum_{k=1}^s\mathrm{tr}\left(E_{0 k}X\right)\right\rvert^2 -\sum_{k=1}^s \left\lvert\mathrm{tr}(E_{0k}X)\right\rvert^2\\
&\approx p\left\lvert\sum_{k=1}^s\alpha_{0 k}\right\rvert^2 -p\sum_{k=1}^s \left\lvert\alpha_{0 k}\right\rvert^2\\
&= p \sum_{k=1}^s \sum_{l\neq k} \alpha_{0 k}\alpha_{0 l}^*.
\end{split}
\label{eq:appExXerrDiff}
\end{align}
As suggested by Preskill~\cite{Preskill97}, this expression has a simple
interpretation in terms of a $2$-D walk composed of $s$ steps of sizes
$\lvert\sqrt{p}\alpha_{0 k}\rvert$.  Equation~(\ref{eq:appExXerrDiff}) is equal to the difference
between the square of the displacement for such a walk and the expectation of the square of the displacement assuming that the walk is random, that is, that stepping forward and backward are equiprobable.  Thus, the expectation of Eq.~(\ref{eq:appExXerrDiff}) vanishes if the sign of
$\alpha_{0 k}$ is random.  Taking a slightly different approach, we can treat the entire expression as the displacement of a $2$-D random
walk composed of $s(s-1)$ steps of sizes $\lvert p \alpha_{0 k}\alpha_{0 l\neq k}\rvert$.  The expectation is again seen to vanish when the sign of $\alpha_{0 k}$ is random, but now it becomes clear that the standard deviation will scale like $s$.  Since $\po{\mathrm{E}X}$ is also proportional to $s$, this implies that the associated stochastic Pauli channel is only really a good substitute when the number of qubits being considered is large.

An identical argument holds for $Y$ and $Z$ errors, showing that, for my purposes, error models for which the sign of the coherent error is random are well approximated by their associated stochastic Pauli channel.  Conveniently, this restriction is preserved under any local relabeling of the Pauli bases and therefore applies equally well to the original error operators.  Examples satisfying the restriction include all stochastic
errors, which have no coherent component, and unitary rotation errors
where under and over rotation are equally likely. Systematic errors, such
as amplitude damping or a bias towards over rotation, are not well
modeled, though, in practice, the local relabeling of the Pauli bases
imposed by gates will randomize these errors somewhat.

\subsection{Implementation\label{sec:implementation}}

In the examples that follow, single-strand error rates were determined for three fault-tolerant procedures.  For each procedure, error rates were calculated for a
universal encoded gate set, $H$, $\CX$, $P$, and $T$ gates, as well as for an idle step
that accounted for the possibility of changing the order of $X$ and $Z$
error correction.  No checks were made on the encoded $T$ gate following
its first error correction, since the remainder of the gate consists of
applying the encoded $P$ gate.  The encoded $P$ gate was assumed
pessimistically to be implemented via a teleportation process akin to that
used for $T$.  For each encoded gate, the maximum was taken over the strand error probabilities at all measurement steps
since in the limit that $n\rightarrow\infty$ encoded failures are caused exclusively by the largest relevant error probability at a measurement
location.

The error probabilities for unencoded gates, measurements, and ancillae were left as
free parameters.  $\po{\Gamma}$, $\pt{\Lambda}{\Xi}$, $\po{M}$,
$\pt{A}{\Gamma}$, and $\pt{B}{\Gamma}$ are used to denote one-qubit,
two-qubit, measurement, $A$-type ancilla, and $B$-type ancilla error
probabilities where $\Gamma$ ranges over the single-qubit Pauli errors and
$\Lambda\Xi$ ranges over the two-qubit Pauli errors.  Note that ancillae
are labeled irrespective of what they encode.  $A$-type ancillae are used
in locations where $Z$ errors are more disruptive than $X$ errors, and
contrariwise for $B$-type ancillae.  In the absence of better information,
I assume that $A$-type ancillae are tested using a homogeneous coupling,
with discard on failure, for first $X$ and then $Z$ errors; the opposite
order is used for $B$-type ancillae.  In this case I approximate the
ancilla error distributions as
\begin{equation}
\begin{split}
\pt{A}{X} &= \pt{X}{Z} + \pt{X}{I} + \pt{I}{X} + \pt{X}{X}\\
\pt{A}{Y} &= \pt{I}{Y} + \pt{X}{Y} \\
\pt{A}{Z} &= \pt{I}{Z} + \pt{X}{Z} \\
\pt{B}{X} &= \pt{X}{I} + \pt{X}{Z} \\
\pt{B}{Y} &= \pt{Y}{I} + \pt{Y}{Z} \\
\pt{B}{Z} &= \pt{I}{Z} + \pt{X}{Z} + \pt{Z}{I} + \pt{Z}{Z}\text{,}
\end{split}\label{eq:ancillaErrors}
\end{equation}
which is (to first order) what one would expect if the only errors on a
verified ancilla were due to undetectable errors on the $\CX$ gates used
to check it.

It should be emphasized that the ancilla error probabilities given by
Eq.~(\ref{eq:ancillaErrors}) are not the only possible choice.  They were
chosen as a good approximation to the residual error following a
verification procedure that discards the state whenever a problem is
indicated. Depending on the purpose of the calculation, it will sometimes
be more appropriate to assign, for example, higher error probabilities
associated with less resource intensive verification or different
probabilities for different kinds of ancillae.

\begin{table}
\capstart
\begin{tabular}{|c|c|}
\hline
\parbox{3.1em}{\centering \rule[.9em]{0em}{0em}Error Model\rule[-.2em]{0em}{0em}} & \parbox{22.5em}{Nonzero Error Probabilities} \\
\hline
\#1 &
\parbox{22.5em}{
$\po{\Gamma}=\frac{p}{4}$\rule{0em}{1.1em},\; $\pt{\Lambda}{\Xi}=\frac{p}{16}$,\; $\po{M}=\frac{p}{2}$,\\
$\pt{A}{X}=\pt{B}{Z}=\frac{p}{4}$,\; $\pt{A}{Y}=\pt{A}{Z}=\pt{B}{X}=\pt{B}{Y}=\frac{p}{8}$\rule[-.5em]{0em}{1.5em} }
\\
\hline
\#2 &
\parbox{22.5em}{
$\po{\Gamma}=\frac{4p}{15}$\rule{0em}{1.1em},\; $\pt{\Lambda}{\Xi}=\frac{p}{15}$,\; $\po{M}=4p$,\\
$\pt{A}{X}=\pt{B}{Z}=\frac{4p}{15}$,\; $\pt{A}{Y}=\pt{A}{Z}=\pt{B}{X}=\pt{B}{Y}=\frac{2p}{15}$\rule[-.5em]{0em}{1.5em} }
\\
\hline
\#3 &
\parbox{22.5em}{
$\pt{\Lambda}{\Xi}=\frac{p}{15}$\rule{0em}{1.1em},\; $\pt{A}{X}=\pt{B}{Z}=\frac{4p}{15}$,\\
$\pt{A}{Y}=\pt{A}{Z}=\pt{B}{X}=\pt{B}{Y}=\frac{2p}{15}$\rule[-.5em]{0em}{1.5em} }
\\
\hline
\#4 &
\parbox{22.5em}{
$\pt{I}{X}=\pt{X}{I}=\pt{I}{Z}=\pt{Z}{I}=\frac{p}{4}$\rule{0em}{1.1em},\\
$\pt{A}{X}=\pt{B}{Z}=\frac{p}{2}$,\; $\pt{A}{Z}=\pt{B}{X}=\frac{p}{4}$\rule[-.5em]{0em}{1.5em} }
\\
\hline
\end{tabular}
\caption{Four reduced error models considered in the text and in Table \ref{tab:thresholdCoefficients}.  $\po{\Gamma}$, $\pt{\Lambda}{\Xi}$, $\po{M}$, $\pt{A}{\Gamma}$, and $\pt{B}{\Gamma}$ represent various one-qubit, two-qubit, measurement, $A$ type ancilla, and $B$ type ancilla error probabilities where $\Gamma\in\{X,Y,Z\}$ and $\Lambda\Xi\in\{I,X,Y,Z\}^{\otimes2}/\{II\}$.  Unspecified probabilities are zero.
\label{tab:errorModels}}
\end{table}

The Mathematica program that I use to calculate encoded error rates retains terms
up to second order in the
base error probabilities, but the results given in the following sections
include only first-order terms.  Second-order terms were found to be
negligible for any plausible choice of error model.  To understand why,
consider a simplified error model in which gates can fail in only a single
way.  Let $p_\mathrm{r}$ be the probability of an individual gate failing,
and let $g_1$ be the number of gates on which a single failure results in
an error at location $\loc{L}$.  Further, let $g_2$ be the number of gates
that might participate in some pair of failures to yield an error at
location $\loc{L}$.  The expected error at location $\loc{L}$ is then
bounded by
\begin{align}
E_{\loc{L}} <& \binom{g_1}{1} p_\mathrm{r} (1-p_\mathrm{r})^{g_2-1} \nonumber \\
&+ \binom{g_2}{2} p_\mathrm{r}^2 (1-p_\mathrm{r})^{g_2-2} + \mathcal{O}(p_\mathrm{r}^3) \\
=& g_1 p_\mathrm{r} -g_1(g_2-1)p_\mathrm{r}^2 + \frac{g_2}{2}(g_2-1) p_\mathrm{r}^2 + \mathcal{O}(p_\mathrm{r}^3). \nonumber
\end{align}
The inequality arises from the fact that not all pairs of failures will
necessarily produce an error at $\loc{L}$.  Even ignoring that, however,
the second-order terms will be negative unless $g_2>2g_1$; negative terms
may safely be neglected since their omission only lowers the threshold.
Among the examples of the following section, the double-coupling Steane
procedure, when applied to error model \#1, has relatively large second-order terms.  Yet the worst location in that procedure corresponds,
roughly, to a single-error situation where $g_1=13$, $g_2=27$, and
$p_\mathrm{r}<.018$, for which the ratio of second to first-order terms is
less than $.02$.  Again, this does not even take into consideration the
fact that many second-order errors will be harmless.

\section{Special Cases\label{sec:examples}}

Having described the operation of my algorithm for calculating
thresholds, I now apply it to three cases of interest.  Two of these are
variants on a fault-tolerant method suggested by Steane~\cite{Steane98},
while the third case is a fault-tolerant telecorrection procedure of the
type proposed by Knill~\cite{Knill04}.

\begin{turnpage}
\begin{table*}[p]
\capstart
\begin{tabular}{|c|c|c|}
\hline
& Gate & Maximal Single-strand Error Probability\\
\hline
& \text{None} & \raisebox{-1.5em}{\rule{0em}{3.5em}}\parbox{21.3cm}{$2 \pt{A}{Y} + 2 \pt{B}{Y} + \pt{I}{Y} + 2 p_{M} + \pt{X}{Y} + \pt{X}{Z} + \pt{Y}{I} + 3 \pt{Y}{X} + 3 \pt{Y}{Y} + \pt{Y}{Z} + 4 \pt{Z}{X} + 3 \pt{Z}{Y} + \max(2 \pt{A}{Z} + 2 \pt{B}{Z} + \pt{I}{Z} + 2 \po{X} + 2 \po{Y} + 3 \pt{Y}{I} + \pt{Y}{X} + 2 \pt{Y}{Z} + 4 \pt{Z}{I} + 3 \pt{Z}{Z},2 \pt{A}{Y} + 2 \pt{B}{Y} + \pt{I}{Y} + 2 \po{M} + \pt{X}{Y} + \pt{X}{Z} + \pt{Y}{I} + 3 \pt{Y}{X} + 3 \pt{Y}{Y} + \pt{Y}{Z} + 4 \pt{Z}{X} + 3 \pt{Z}{Y})$} \\
\cline{2-3}
& $H$ & \raisebox{-1.5em}{\rule{0em}{3.5em}}\parbox{21.3cm}{$\pt{A}{X} + 2 \pt{A}{Y} + \pt{A}{Z} + \pt{B}{Y} + \pt{B}{Z} + \pt{I}{X} + \pt{I}{Y} + 2 \po{M} + \po{X} + \pt{X}{Z} + 2 \po{Y} + 2 \pt{Y}{I} + 2 \pt{Y}{X} + 2 \pt{Y}{Y} + \pt{Y}{Z} + 2 \pt{Z}{I} + 3 \pt{Z}{X} + 2 \pt{Z}{Y} + \pt{Z}{Z} + \max(\pt{X}{I} + \pt{Y}{I} + 2 \pt{Y}{Z} + \po{Z} + \pt{Z}{Y} + \pt{Z}{Z},\pt{I}{Y} + \pt{I}{Z} + \po{X} + \pt{X}{X} + 2 \pt{X}{Y} + \pt{Y}{X})$} \\
\cline{2-3}
& $\CX$ & \raisebox{-1.5em}{\rule{0em}{3.5em}}\parbox{21.3cm}{$\pt{A}{Y} + \pt{B}{Y} + 2 \pt{I}{Y} + 3 p_{M} + 2 \pt{X}{Y} + 2 \pt{X}{Z} + 2 \pt{Y}{I} + 3 \pt{Y}{X} + 3 \pt{Y}{Y} + 2 \pt{Y}{Z} + 5 \pt{Z}{X} + 3 \pt{Z}{Y} + \max(\pt{A}{Z} + 2 \pt{B}{Y} + 3 \pt{B}{Z} + 2 \pt{I}{Z} + 3 \po{X} + 3 \po{Y} + 3 \pt{Y}{I} + 2 \pt{Y}{X} + \pt{Y}{Z} + 5 \pt{Z}{I}, 3 \pt{A}{X} + 2 \pt{A}{Z} + \pt{B}{X} + 5 \pt{I}{X} + 3 \pt{I}{Y} + 2 \pt{X}{I} + 3 \pt{X}{X} + \pt{X}{Y} + 2\pt{Z}{Y})$} \\
\cline{2-3}
\rule{.9em}{0em}\turnbox{90}{Single-coupling Steane}\rule{.3em}{0em} & $T$, $P$ & \raisebox{-1.5em}{\rule{0em}{3.5em}}\parbox{21.3cm}{$\pt{B}{Y} + \po{M} + \pt{Y}{I} + \pt{Y}{X} + \pt{Y}{Y} + \pt{Y}{Z} + \pt{Z}{X} + \pt{Z}{Y} + \max(\pt{A}{X} + \pt{A}{Y} + \pt{B}{X} + 2 \pt{I}{X} + 2 \pt{I}{Y} + \po{M} + \pt{X}{I} + \pt{X}{X} + \pt{X}{Y} + \pt{X}{Z} + \pt{Z}{X} + \pt{Z}{Y},\pt{B}{Y} + 2 \pt{B}{Z} + \po{X} + \po{Y} + \pt{Z}{I} + \pt{Z}{Z})$} \\
\hline
& \text{None} & \raisebox{-1.5em}{\rule{0em}{3.5em}}\parbox{21.3cm}{$\pt{A}{Y} + \pt{B}{Y} + 3 \pt{I}{Y} + \po{M} + 3 \pt{X}{Y} + 3 \pt{X}{Z} + 3 \pt{Y}{I} + 5 \pt{Y}{X} + 7 \pt{Y}{Y} + 3 \pt{Y}{Z} + 5 \pt{Z}{X} + 5 \pt{Z}{Y} + \max(3 \pt{A}{Y} + 4 \pt{A}{Z} + \pt{B}{Z} + 3 \pt{I}{Z} + \po{X} + \po{Y} + 2 \pt{Y}{I} + 4 \pt{Y}{Z} + 5 \pt{Z}{I} + 2 \pt{Z}{Y} + 7 \pt{Z}{Z},\pt{A}{X} + 4 \pt{B}{X} + 3 \pt{B}{Y} + 5 \pt{I}{X} + 2 \pt{I}{Y} + 3 \pt{X}{I} + 7 \pt{X}{X} + 4 \pt{X}{Y} + 2 \pt{Y}{X})$} \\
\cline{2-3}
& $H$ & \raisebox{-1.5em}{\rule{0em}{3.5em}}\parbox{21.3cm}{$2 \pt{A}{Y} + 2 \pt{A}{Z} + \pt{I}{Y} + \pt{I}{Z} + \po{M} + \po{X} + \pt{X}{I} + \pt{X}{X} + 2 \pt{X}{Y} + 3 \pt{X}{Z} + \po{Y} + 3 \pt{Y}{I} + 4 \pt{Y}{X} + 5 \pt{Y}{Y} + 5 \pt{Y}{Z} + 2 \pt{Z}{I} + 3 \pt{Z}{X} + 4 \pt{Z}{Y} + 3 \pt{Z}{Z} + \max(\pt{A}{X} + \pt{A}{Y} + \pt{I}{X} + 2 \pt{I}{Y} + \pt{I}{Z} + \pt{X}{X} + 2 \pt{X}{Y},\pt{B}{Y} + \pt{B}{Z} + \pt{X}{I} + \po{Y} + 2 \pt{Y}{I} + 2 \pt{Y}{Z} + \po{Z} + \pt{Z}{I} + \pt{Z}{Z})$} \\
\cline{2-3}
& $\CX$ & \raisebox{-1.5em}{\rule{0em}{3.5em}}\parbox{21.3cm}{$\pt{A}{Y} + \pt{B}{Y} + 4 \pt{I}{Y} + \po{M} + 5 \pt{X}{Y} + 5 \pt{X}{Z} + 4 \pt{Y}{I} + 4 \pt{Y}{X} + 7 \pt{Y}{Y} + 5 \pt{Y}{Z} + 4 \pt{Z}{X} + 4 \pt{Z}{Y} + \max(\pt{A}{Y} + 2 \pt{A}{Z} + \pt{B}{Z} + \pt{I}{Y} + 5 \pt{I}{Z} + \po{X} + \po{Y} + 2 \pt{Y}{Z} + 4 \pt{Z}{I} + 3 \pt{Z}{Y} + 7 \pt{Z}{Z}, \pt{A}{X} + 2 \pt{B}{X} + \pt{B}{Y} + 4 \pt{I}{X} + 5 \pt{X}{I} + 7 \pt{X}{X} + 2 \pt{X}{Y} + \pt{Y}{I} + 3 \pt{Y}{X})$} \\
\cline{2-3}
\rule{.9em}{0em}\turnbox{90}{Double-coupling Steane}\rule{.3em}{0em} & $T$, $P$ & \raisebox{-1.5em}{\rule{0em}{3.5em}}\parbox{21.3cm}{$2 \pt{B}{Y} + \pt{I}{Y} + \po{M} + \pt{X}{Y} + \pt{Y}{X} + 2 \pt{Y}{Y} + \pt{Z}{X} + 2 \pt{Z}{Y} + \max(2 \pt{B}{X} + 2 \pt{I}{X} + \pt{I}{Y} + 2 \pt{X}{I} + 3 \pt{X}{X} + 2 \pt{X}{Y} + 2 \pt{X}{Z} + 2 \pt{Y}{I} + 2 \pt{Y}{X} + \pt{Y}{Y} + 2 \pt{Y}{Z} + \pt{Z}{X},2 \pt{B}{Z} + \pt{I}{Z} + \po{X} + \pt{X}{Z} + \po{Y} + \pt{Y}{I} + 2 \pt{Y}{Z} + \pt{Z}{I} + 2 \pt{Z}{Z},\pt{A}{X} + \pt{A}{Y} + 3 \pt{B}{X} + \pt{B}{Y} + 3 \pt{I}{X} + 2 \pt{I}{Y} + 3 \pt{X}{X} + 2 \pt{X}{Y} + 2 \pt{Y}{X} + \pt{Y}{Y} + 2 \pt{Z}{X} + \pt{Z}{Y})$} \\
\hline
& \parbox{3em}{None, $T$, $P$} & \raisebox{-1em}{\rule{0em}{2.5em}}\parbox{21.6cm}{$\pt{A}{Y} + \pt{B}{Y} + \po{M} + \pt{Y}{X} + \pt{Y}{Y} + \pt{Z}{X} + \pt{Z}{Y} + \max(\pt{A}{Z} + \pt{B}{Z} + \po{X} + \po{Y} + \pt{Y}{I} + \pt{Y}{Z} + \pt{Z}{I} + \pt{Z}{Z},\pt{A}{X} + \pt{B}{X} + \pt{I}{X} + \pt{I}{Y} + \pt{X}{X} + \pt{X}{Y})$} \\
\cline{2-3}
& $H$ & \raisebox{-1em}{\rule{0em}{2.5em}}\parbox{21.3cm}{$\pt{B}{X} + 2 \pt{B}{Y} + \pt{B}{Z} + \po{M} + \po{X} + \po{Y} + \pt{Y}{X} + \pt{Y}{Y} + \pt{Z}{X} + \pt{Z}{Y} + \min(\pt{I}{X} + \pt{I}{Y} + \pt{X}{X} + \pt{X}{Y},\po{Y} + \pt{Y}{I} + \pt{Y}{Z} + \po{Z} + \pt{Z}{I} + \pt{Z}{Z})$} \\
\cline{2-3}
\rule{.9em}{0em}\turnbox{90}{\ \ Knill}\rule{.3em}{0em} & $\CX$ & \raisebox{-2em}{\rule{0em}{4.5em}}\parbox{21.3cm}{$\pt{A}{Y} + \pt{B}{Y} + \po{M} + \pt{Y}{X} + 2 \pt{Y}{Y} + \pt{Z}{X} + \pt{Z}{Y} + \max(\pt{A}{Y} + 2 \pt{A}{Z} + \pt{B}{Z} + \po{X} + \po{Y} + 2 \pt{Y}{I} + \pt{Y}{X} + 2 \pt{Y}{Z} + 2 \pt{Z}{I} + \pt{Z}{X} + \pt{Z}{Y} + 2 \pt{Z}{Z},\pt{A}{X} + \pt{B}{X} + \pt{I}{X} + \pt{I}{Y} + \pt{X}{I} + 2 \pt{X}{X} + 2 \pt{X}{Y} + \pt{X}{Z} + \pt{Y}{I} + \pt{Y}{X} + \pt{Y}{Z},\pt{A}{X} + 2 \pt{B}{X} + \pt{B}{Y} + 2 \pt{I}{X} + 2 \pt{I}{Y} + 2 \pt{X}{X} + 2 \pt{X}{Y} + \pt{Y}{X} + \pt{Z}{X} + \pt{Z}{Y},\pt{A}{Z} + \pt{B}{Z} + \pt{I}{Y} + \pt{I}{Z} + \po{X} + \pt{X}{Y} + \pt{X}{Z} + \po{Y} + \pt{Y}{I} + 2 \pt{Y}{Z} + \pt{Z}{I} + \pt{Z}{Y} + 2 \pt{Z}{Z})$} \\
\hline
\end{tabular}
\caption{Maximum error probabilities for a single strand of various encoded gates for a selection of fault-tolerant procedures.  The subscripted $p$'s refer to probabilities of error of various unencoded operations with $X$, $Y$, $Z$ denoting Pauli errors, $M$ measurement errors, and $A$ and $B$ ancilla errors.  Thus $\po{X}$, $\pt{I}{Z}$, $\pt{Z}{I}$, $\po{M}$, and $\pt{A}{Y}$ denote the probabilities of a single-qubit-gate $X$ error, a $Z$ error on the target of a $\protect \CX$, a $Z$ error on the control of a $\protect \CX$, a measurement error, and a $Y$ error on an $A$ type ancilla.  A procedure is below threshold when the maximum error probability for all gates is less than $\tau$, the fraction of errors corrected asymptotically. \label{tab:encodedErrorRates}}
\end{table*}
\end{turnpage}

The unrefined output of this endeavor is the set of maximum
strand error probabilities listed in Table~\ref{tab:encodedErrorRates}.
This table specifies a kind of high-dimensional threshold surface in the space of generic
stochastic error models.  An error channel is below the threshold for a particular procedure
whenever the maximal strand error probabilities for that procedure are lower than the fraction
of errors that are correctable asymptotically.  For the purpose of illustration, however, it is more useful to consider
less complicated error models.  Table~\ref{tab:errorModels} defines four
reduced error models in terms of the generic stochastic error model.  Since these
reduced error models have only a single free parameter, their threshold surfaces are
simply numbers.  Table~\ref{tab:thresholdCoefficients} lists thresholds for three
procedures and four reduced error models in terms of $\tau$, the asymptotic
correctable error fraction.

The following subsections provide supplemental information specific to
each procedure, including qualitative reviews of the procedures, circuit
diagrams for encoded gates, and commentary on the thresholds given in
Table~\ref{tab:thresholdCoefficients}.

\begin{table}
\capstart
\begin{tabular}{|c|r@{.}l|r@{.}l|r@{.}l|r@{.}l|}
\hline
\multicolumn{9}{|c|}{Thresholds for Homogeneous Ancillae ($\tau$)\rule[-.8em]{0em}{2.2em}} \\
\hline
\setlength{\unitlength}{1em}
\begin{picture}(12,1.79)
\put(-.53,1.78){\line(6,-1){12.71}}
\put(-.1,.02){Procedure}
\put(6.4,.82){Error Model}
\end{picture}
& \multicolumn{2}{|c|}{\raisebox{.4em}{\#1}} & \multicolumn{2}{|c|}{\raisebox{.4em}{\#2}} & \multicolumn{2}{|c|}{\raisebox{.4em}{\#3}} & \multicolumn{2}{|c|}{\raisebox{.4em}{\#4}} \\
\hline
Single-coupling Steane\rule[-.5em]{0em}{1.6em} & 0&15 & 0&06 & 0&24 & 0&29 \\
\hline
Double-coupling Steane\rule[-.5em]{0em}{1.6em} & 0&16 & 0&10 & 0&18 & 0&29 \\
\hline
Knill\rule[-.5em]{0em}{1.6em} & \:0&35\: & \:0&15\: & \:0&50\: & \:0&67\: \\
\hline
\end{tabular}
\caption{Threshold for ancillae with homogeneous errors given in units of $\tau$, the correctable error fraction, for the three procedures and four error models I consider as examples.  These thresholds were obtained by substituting the parameters given in Table~\ref{tab:errorModels} into Table~\ref{tab:encodedErrorRates} and requiring that single-strand error probabilities not exceed $\tau$. \label{tab:thresholdCoefficients} }
\end{table}

\subsection{Steane's Method}
Steane's method is characterized by the extraction of error information through the coupling of encoded data to ancillae prepared in simple logical states.  Using an ancilla prepared in the logical state $\ket{\bar{0}}$ ($\ket{\bar{+}}$) it is possible via a single transversal two-qubit operation to extract information about the location of $Z$ ($X$) errors on all qubits in a data block.

Typical instantiations of Steane's method employ multiple extractions to guard against errors made during the coupling and measurement process.
Often \cite{Steane03,Reichardt04,Zalka96} the number of extractions performed is conditional on their output.  I deviate from this rule by demanding a fixed number of couplings.  This a sensible choice for my analysis since, up to rearrangement of qubits, the output of an extraction becomes deterministic as the number of qubits in an encoding approaches infinity.  Moreover, if $p$ is the probability of an error occurring, requiring the sequential agreement of $j$ extractions reduces the probability of misdiagnosing an error on a particular line to order $p^j$.  Since I ultimately retain only first-order terms I need only consider single and double-coupling Steane procedures.

\subsubsection{Single-coupling Steane Procedure}

\begin{figure}[tbp]
\capstart
\begin{tabular*}{26.5em}{@{\extracolsep{\fill}}lll}
\Qcircuit @R=.3em @C=.4em @!R {
& \text{a)} & & & & & & & & & & {\ F_X} \\
& & & & & & & & {A_A} & & & \targ & \meter \\
& & & & & \qw & \push{\rule[-.4em]{0em}{.4em}} \qw & \qw & \qw & \qw & \qw & \ctrl{-1} & \qw & \qw & \qw
\gategroup{3}{7}{2}{13}{.6em}{--}
}
&
\Qcircuit @R=.3em @C=.4em @!R {
& \text{b)} & & & & & & & & & & & {\!\!\!\!F_Z} \\
& & & & & & & & {A_B} & & & \ctrl{1} & \gate{H} & \meter \\
& & & & & \qw & \push{\rule[-.6em]{0em}{.8em}} \qw & \qw & \qw & \qw & \qw & \targ & \qw & \qw & \qw & \qw
\gategroup{3}{7}{2}{14}{.6em}{--}
}
\\
\\
c)
\end{tabular*}
\\
\begin{tabular*}{26.5em}{@{\extracolsep{\fill}}|c|c|}
\hline
\:Gate\: & Circuit \\
\hline
\raisebox{.1em}{None} &
%\ \ \ \
\rule{2.4em}{0em}
\raisebox{.45em}{\rule[-1em]{0em}{2em}
\Qcircuit @R=.4em @C=.8em {
& \gate{F_{Z}} &  \gate{F_{X}} &  \gate{F_{X}} &  \gate{F_{Z}} & \gate{F_{Z}} &  \gate{F_{X}} & \qw
}
}
%\ \ \ \
\rule{2.4em}{0em}
\\
\hline
\raisebox{.1em}{$H$} &
\raisebox{.45em}{\rule[-1em]{0em}{2em}
\Qcircuit @R=.4em @C=.8em {
& \gate{F_{Z}} &  \gate{F_{X}} &  \gate{H} &  \gate{F_{X}} &  \gate{F_{Z}} & \qw
}
}
\\
\hline
\raisebox{-.4em}{$\CX$} &
\raisebox{.9em}{\rule[-2.8em]{0em}{3.8em}
\Qcircuit @R=.4em @C=.8em {
& \gate{F_{Z}} &  \gate{F_{X}} &  \ctrl{1} &  \gate{F_{Z}} &  \gate{F_{X}} & \qw \\
& \gate{F_{X}} &  \gate{F_{Z}} &  \targ &  \gate{F_{X}} &  \gate{F_{Z}} & \qw
}
}
\\
\hline
\raisebox{-.2em}{$P$, $T$} &
\raisebox{.9em}{\rule[-2.8em]{0em}{3.9em}
\Qcircuit @R=.4em @C=.8em {
& \gate{F_{X}} &  \gate{F_{Z}} &  \targ &  \meter \\
& & \lstick{A_B} & \ctrl{-1} & \gate{F_Z} & \qw
}
}
\\
\hline
\end{tabular*}
\caption{Encoded circuits for the single-coupling Steane procedure.  Handling of the data is minimized at the cost of syndrome verification.  Parts a and b display the circuits for finding $X$ and $Z$ errors, respectively.  Part c lists the circuits analyzed to determine the encoded error rates for this example.\label{fig:singleSteane}}
\end{figure}
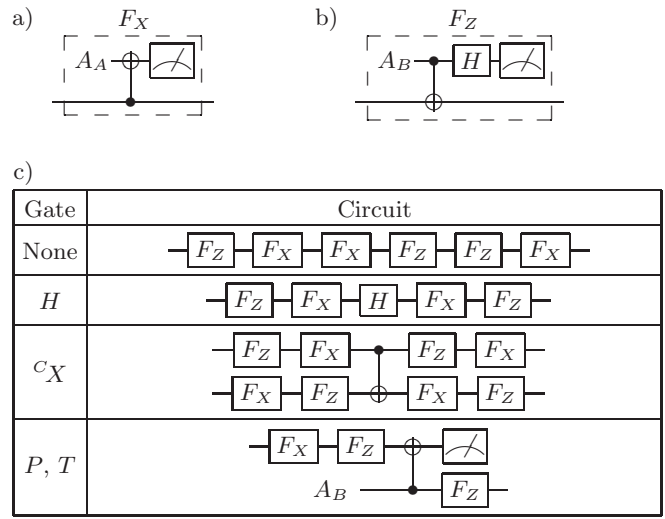

Two-qubit gates require a controlled interaction between two otherwise
isolated quantum systems.  Consequently, they are often the most error
prone gates in a universal set.  In such cases, the factor limiting the
probability of successful error correction may be the number of times that
two-qubit gates must be applied to the data in order to reliably diagnose
errors.  For Steane's method, this interaction is minimized by coupling to
the data once per $X$ correction and once per $Z$ correction, as shown in
Fig.~\ref{fig:singleSteane}.

Table~\ref{tab:thresholdCoefficients} shows that, relative to the other
procedures considered, the single-coupling Steane procedure performs most
strongly for error model \#3.  This is in line with our expectations since
model \#3 includes only two-qubit gate errors and the resultant ancilla
errors.  Surprisingly, it also does rather well overall, suffering in
comparison to the double-coupling Steane procedure only for error model
\#2 where measurement errors dominate.  The single-coupling Steane
procedure lacks a means of syndrome verification, so any errors in
syndrome measurement are transferred directly to the data. Nevertheless,
my results demonstrate that moderate single-qubit and measurement error
probabilities can be tolerated when high quality ancillae are available.

\subsubsection{Double-coupling Steane Procedure}

\begin{figure}[btp]
\capstart
\begin{tabular*}{26.5em}{@{\extracolsep{\fill}}ll}
\Qcircuit @R=.3em @C=.4em @!R {
& \text{a)} & & & & & & & {F^\prime_X} \\
& & & & & & & & & {A_A\;\;\;\;\;\;} & \targ &  \meter \\
& & & & {A_A} & & & \targ & \meter \\
& & \qw & \push{\rule[-.6em]{0em}{.8em}} \qw & \qw & \qw & \qw & \ctrl{-1} & \qw & \qw & \ctrl{-2} & \qw & \qw & \qw
\gategroup{4}{4}{2}{12}{.6em}{--}
}
&
\Qcircuit @R=.3em @C=.4em @!R {
& \text{b)} & & & & & & & & {F^\prime_Z} \\
& & & & & & & & & & {A_B\;\;\;\;\;\;} & \ctrl{2} & \gate{H} &  \meter \\
& & & & {A_B} & & & \ctrl{1} & \gate{H} & \meter \\
& & \qw & \push{\rule[-.6em]{0em}{.8em}} \qw & \qw & \qw & \qw & \targ & \qw & \qw & \qw & \targ & \qw & \qw & \qw & \qw
\gategroup{4}{4}{2}{14}{.6em}{--}
}
\\
\\
c)
\end{tabular*}
\\
\begin{tabular*}{26.5em}{@{\extracolsep{\fill}}|c|c|}
\hline
\:Gate\: & Circuit \\
\hline
\raisebox{.2em}{None} &
\rule{2.4em}{0em}
\raisebox{.45em}{\rule[-1.1em]{0em}{2.2em}
\Qcircuit @R=.4em @C=.8em {
& \gate{F^\prime_Z} &  \gate{F^\prime_X} &  \gate{F^\prime_X} &  \gate{F^\prime_Z} & \gate{F^\prime_Z} &  \gate{F^\prime_X} & \qw
}
}
\rule{2.4em}{0em}
\\
\hline
\raisebox{.2em}{$H$} &
\raisebox{.45em}{\rule[-1.1em]{0em}{2.2em}
\Qcircuit @R=.4em @C=.8em {
& \gate{F^\prime_Z} &  \gate{F^\prime_X} &  \gate{H} &  \gate{F^\prime_X} &  \gate{F^\prime_Z} & \qw
}
}
\\
\hline
\raisebox{-.4em}{$\CX$} &
\raisebox{.9em}{\rule[-3.1em]{0em}{4.3em}
\Qcircuit @R=.4em @C=.8em {
& \gate{F^\prime_Z} &  \gate{F^\prime_X} &  \ctrl{1} &  \gate{F^\prime_Z} &  \gate{F^\prime_X} & \qw \\
& \gate{F^\prime_X} &  \gate{F^\prime_Z} &  \targ &  \gate{F^\prime_X} &  \gate{F^\prime_Z} & \qw
}
}
\\
\hline
\raisebox{-.2em}{$P$, $T$} &
\raisebox{.9em}{\rule[-3.1em]{0em}{4.3em}
\Qcircuit @R=.4em @C=.8em {
& \gate{F^\prime_X} &  \gate{F^\prime_Z} &  \targ &  \meter \\
& & \lstick{A_B} & \ctrl{-1} & \gate{F^\prime_Z} & \gate{F_X} & \qw
}
}
\\
\hline
\end{tabular*}
\caption{Encoded circuits for the double-coupling Steane procedure.  Syndrome information is extracted twice, and qubits implicated both times are presumed to be in error.  Parts a and b display the circuits for finding $X$ and $Z$ errors, respectively.  Part c lists the circuits analyzed to determine the encoded error rates for this example.\label{fig:doubleSteane}}
\end{figure}
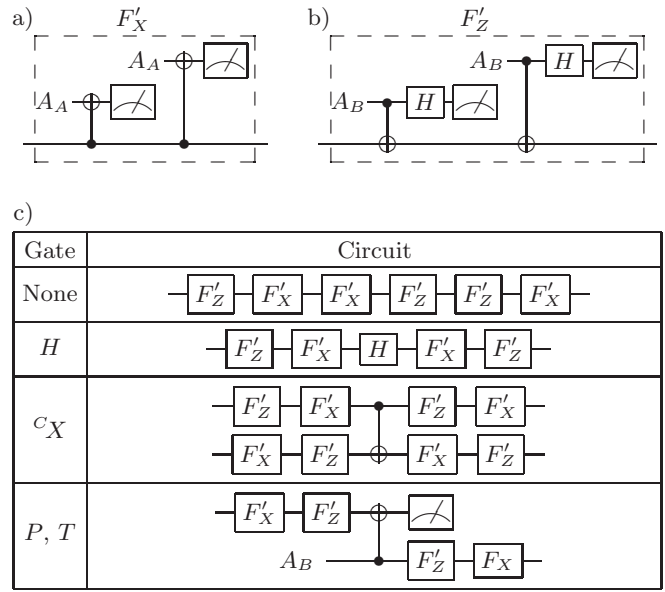

When two-qubit gates are relatively reliable, the damage done during the
extraction of error information can be limited by preparing ancillae such
that they include few errors capable of propagating to the data.   Under
these circumstances, it is often advantageous to verify error diagnoses by
coupling to the data more than once, as shown in
Fig.~\ref{fig:doubleSteane}.

For the double-coupling Steane procedure, error model \#1 is especially
interesting because it was chosen in imitation of the error model used by
Reichardt~\cite{Reichardt04} in his numerical estimation of the threshold
for a Steane style procedure on a 49 qubit code.  My threshold of $0.90\%$
for asymptotic minimum distance decoding is quite close to his value of
roughly $0.88\%$.  The extraordinary agreement of these two estimates is a
coincidence, as can be seen from my discussion of finite codes in
\S\ref{sec:finiteCodes}, but their rough equivalence illustrates the value
of my idealized algorithm for approximating the encoded error rates used in
threshold estimation.  Implementing Steane's method with 49 qubit ancillae
prepared via a liberal discard policy yields roughly the same
threshold as implementing it with very large ancillae prepared such that
errors due to the two-qubit gates used for verification dominate.

As expected, relative to the other two procedures, the double-coupling
Steane procedure performs most favorably for error model \#2.  Somewhat
surprisingly, however, it still underperforms the Knill procedure.  The
reason for this is most easily understood by considering the limiting case
in which only measurement errors occur.  In the absence of any other
source of error, measurement errors have no effect on either the double-coupling Steane or the Knill procedure until their probability exceeds
$\tau$; beyond that point both procedures fail with certainty.  Thus, the
two procedures cope with measurement errors equally well, but the Knill
procedure handles other kinds of gate errors more effectively.

Error models \#3 and \#4 demonstrate small gains in the threshold that can
result when two-qubit gate errors have some underlying structure.  Model
\#3 is a pure two-qubit-gate depolarizing error model (which includes the
associated ancilla errors) while \#4 is a model in which two-qubit gates
malfunction by producing either an $X$ or a $Z$ error on either the
control or the target.  Given the highly restricted form of error model
\#4 it is discouraging that the threshold increases by less than a factor
of two over that of error model \#3.  Nonetheless, the possibility of
larger gains in a procedure designed to take advantage of some particular
error distribution is not ruled out.

\subsection{Knill's Method}

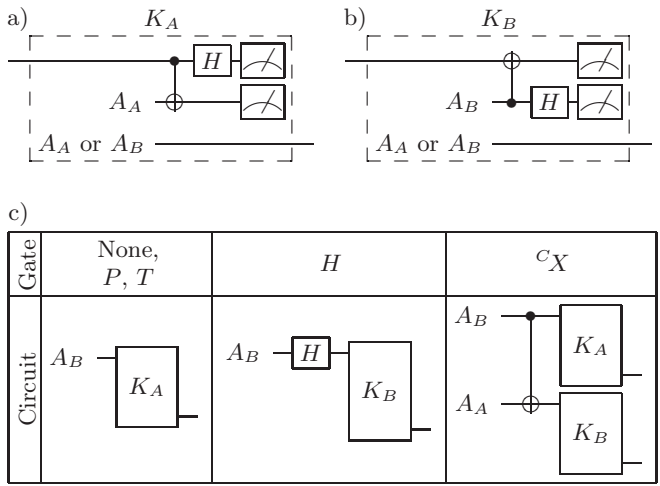
\begin{figure}[tbp]
\capstart
\begin{tabular*}{26.5em}{@{\extracolsep{\fill}}ll}
\Qcircuit @R=.3em @C=.4em @!R {
& \text{a)} & & & & {\!\!\!\!\!\!K_A}\\
& \qw & \qw & \qw & \qw & \ctrl{1} & \gate{H} & \meter \\
& & & & \lstick{A_A} & \targ & \qw & \meter \\
& & & \push{A_A\text{ or } A_B} & & \qw & \qw & \qw & \qw & \qw & \qw
\gategroup{4}{4}{2}{8}{.6em}{--}
}
&
\Qcircuit @R=.3em @C=.4em @!R {
& \text{b)} & & & & {\!\!\!\!\!\!K_B}\\
& \qw & \qw & \qw & \qw & \targ & \qw & \meter \\
& & & & \lstick{A_B} & \ctrl{-1} & \gate{H} & \meter \\
& & & \push{A_A\text{ or } A_B} & & \qw & \qw & \qw & \qw & \qw & \qw
\gategroup{4}{4}{2}{8}{.6em}{--}
}
\\
\\
c)
\end{tabular*}
\\
\begin{tabular*}{26.5em}{@{\extracolsep{\fill}}|c|c|c|c|}
\hline
\raisebox{-.8em}{\rule{.9em}{0em}\turnbox{90}{Gate}\rule[-.2em]{0em}{2.6em}} & \parbox{3em}{None, $P$, $T$} & $H$ & $\CX$
\\
\hline
\raisebox{-3em}{\rule{.9em}{0em}\turnbox{90}{Circuit}} &
\raisebox{-.2em}{
\Qcircuit @R=.3em @C=.8em @!R {
& & \lstick{A_B} & \multigate{2}{K_A}\\
& & & \pureghost{K_A}\\
& & & \pureghost{K_A} & \qw
}
}\rule{.4em}{0em}
&
\rule{.2em}{0em}
\Qcircuit @R=.3em @C=.8em @!R {
& & \lstick{A_B} & \gate{H} & \multigate{2}{K_B}\\
& & & & \pureghost{K_B}\\
& & & & \pureghost{K_B} & \qw
}\rule{.4em}{0em}
&
\raisebox{1.5em}{
\Qcircuit @R=.3em @C=.8em @!R {
& & \lstick{A_B} & \ctrl{3} & \multigate{2}{K_A}\\
& & & & \pureghost{K_A}\\
& & & & \pureghost{K_A} & \qw \\
& & \lstick{A_A} & \targ & \multigate{2}{K_B}\\
& & & & \pureghost{K_B}\\
& & & & \pureghost{K_B} & \qw
}
}\rule[-5.3em]{0em}{7.6em}\rule{.4em}{0em}
\\
\hline
\end{tabular*}
\caption{Encoded circuits for the Knill procedure.  Error correction is performed by teleporting the data, minus the errors, using an entangled two-logical-qubit ancilla; the precise location of errors is unimportant so long as the result of the logical-qubit measurement is correctly decoded.  Parts a and b display the circuits for correcting errors when the output of the previous step was dominated by $Z$ and $X$ errors respectively.  Part c lists the circuits analyzed to determine the encoded error rates for this example.  Ancilla error distributions are used as the input to these circuits since that is the only remaining source of error after a successful teleportation correction.\label{fig:Knill}}
\end{figure}

Knill's method for fault-tolerant quantum computation~\cite{Knill05} utilizes telecorrection (see Fig.~\ref{fig:Knill}).  Telecorrection, error correction by teleportation, requires the application of even fewer two-qubit gates to the data undergoing correction than a single-coupling Steane error correction.  The cost of this innovation is a greater reliance on the ability to produce high quality ancillae.  In effect, Knill's method exchanges the error distribution of an encoded data block for that of one half of an ancilla prepared in a logical Bell state.  Failure occurs only when too many errors are present at the time of measurement to correctly identify the encoded Pauli operator needed to complete the teleportation.  Conveniently, encoded single-qubit gates can be accomplished by performing the teleportation using a logical Bell state prepared with the desired gate already applied to one half.

For my implementation of Knill's method, error model \#4 achieves the highest threshold, though physical systems displaying this sort of error seem unlikely.  Error model \#1 provides another check of my algorithm, since its parameters are also roughly those used by Knill \cite{Knill04} in a paper on telecorrection.  Setting $\tau$ to $11\%$ for the channel capacity for CSS codes, I find that the threshold for this model is $3.9\%$ compared to Knill's estimate of $3\%$ and his extrapolation of up to $5\%$.  The approximate agreement between these values is satisfying, though an exact match is not expected since Knill assumes that errors on up to $19\%$ of the qubits can be corrected, an assumption that derives from bounds on the channel capacity for general quantum codes~\cite{DiVincenzo98}.

Of course the most striking aspect of
Table~\ref{tab:thresholdCoefficients} is that the Knill procedure yields a
higher threshold for every error model.  As with the single-coupling
Steane case, this derives partly from my assumptions regarding ancillae.
In particular, Steane's method was designed to utilize ancillae for
which either correlated $X$ or correlated $Z$ errors could be minimized,
but not both, a situation certain to favor his approach.  A second but
lesser objection can be made that I set the ancilla error probabilities
equal for all gates and all methods, ignoring the fact that some methods,
such as Knill's, and some gates, such as the $T$ gate, will require more
complex ancillae which may in turn be more error prone.  Substantially
more detailed ancilla information would be needed to evaluate the
importance of this effect, but the overall character of my results is
unlikely to change since that would entail in excess of a two-fold
increase in the error probabilities for logical two-qubit ancillae over
those for ancillae prepared in a single-qubit logical state.  Thus, so
long as resource considerations do not limit our ability to discard
suspect ancillae, and therefore to make very high quality ancillae,
Knill's method will provide the highest thresholds.

\section{Finite Codes\label{sec:finiteCodes}}
Prior to taking the limit $n\rightarrow\infty$, the expression for the
probability of an encoded error at a location $\loc{L}$ was
\begin{align}
E_{\mathrm{f}}(p_\loc{L}) = \sum_{i=t+1}^n\binom{n}{i}p_\loc{L}^i(1-p_\loc{L})^{n-i},\label{eq:pfail}
\end{align}
where $t$ is the number of correctable errors and $p_\loc{L}$ is the
probability of a relevant error on a single qubit at the location in
question.  Using this expression, the programme of \S\ref{sec:homogeneous}
can be implemented for finite $n$.  In doing so, however, the simplicity
of the algorithm suffers somewhat, and its interpretation as an idealized
threshold bound is completely lost.  Fundamentally, the complications that
arise are all due to the fact that the success or failure of various
portions of an encoded gate are no longer deterministic.  This section
explains how to deal with the associated difficulties and concludes with a
brief demonstration of the algorithm for a $[[49,1,9]]$ code.

In the examples of \S\ref{sec:examples}, I establish a background error
rate by performing an initial error correction, but for finite $n$ this
initialization is not guaranteed to succeed.  Though the failure of the
initial error correction is properly assigned to the previous encoded
gate, the residual errors will differ dramatically depending on whether it
occurred.  This presents no problem when only a single level of encoding
is employed since any encoded failure is considered a failure of the
computation.  In concatenated coding schemes, however, failed encoded
qubits are corrected at higher levels of encoding.  Their continued use is
problematic since an encoded gate failure may be correlated with
subsequent encoded failures.  Nevertheless, I recommend calculating the
encoded error rate for finite codes using the assumption that the
initialization did not fail, a choice that requires no modification to the
case for large $n$.

Likewise, calculation of the single line error rate $p_\loc{L}$ proceeds
without modification.  For finite codes, however, the maximum tolerable
single line error rate becomes a nontrivial function of the encoded error
rate that we wish to achieve.  The probability of an unrecoverable error
never goes to zero, so it is necessary to perform the summation in
Eq.~(\ref{eq:pfail}) to determine the portion of the encoded error rate
due to any particular location.

The possibility of failure must be considered at many points in the
circuit since statistical fluctuations will produce unrecoverable errors
at a variety of locations.  Typically, encoded failure probabilities at
various locations will be strongly correlated, but the exact nature of
these correlations is difficult to predict.  Thus, the best I can do
is to bound the encoded failure probability,
\begin{align}
\max_{\loc{L}\in\loc{S}} E_{\mathrm{f}}(p_\loc{L}) \leq \left\{
\parbox{5.3em}{\centering Encoded error rate}\rule{0em}{1.3em}
\right\} \leq \sum_{\loc{L}\in\loc{S}} E_{\mathrm{f}}(p_\loc{L})\label{eq:encodedRange}
\end{align}
where $\loc{S}$ ranges over the locations of every post-initialization output, that is, syndrome measurements and the final state of the data with regard to both $X$ and $Z$ errors.

To clarify the changes outlined above, consider the example of the double-coupling Steane procedure implemented using a $[[49,1,9]]$ quantum code and subject to the error channel defined by error model \#1.  For the encoded $\CX$ gate, the set of single line error probabilities corresponding to $X$ errors at the eight locations of post-initialization syndrome measurement and $X$ and $Z$ errors at the two output locations of the data is
\begin{align}
\begin{split}
\{p_\loc{S}\}=\left\{\frac{47p}{8},\frac{43p}{8},\frac{43p}{8},\right.&\frac{41p}{8},\frac{39p}{8},\frac{33p}{8},\frac{37p}{8},\\
&\left.\frac{31p}{8},\frac{9p}{4},\frac{9p}{4},\frac{3p}{4},\frac{3p}{4}\right\}.
\end{split}
\end{align}
Solving Eq.~(\ref{eq:encodedRange}) subject to the restriction that the encoded error rate is exactly $p$ yields solutions in the range
\begin{align}
.0036 \geq p \geq .0023.
\end{align}
Repeating this process for each of the other encoded gates and taking the minimum over the upper and lower bounds produces a threshold of
\begin{align}
.0034 \geq p_{th} \geq .0023
\end{align}
where, of course, the caveats discussed in \S\ref{subsec:thresholds} regarding concatenated threshold estimates all apply.  This example provides a particularly apt comparison to Reichardt's threshold estimate for the $[[49,1,9]]$ code~\cite{Reichardt04}.  The threshold calculated here is roughly a third of that estimated by Reichardt.  The difference presumably springs from the superiority of his rule for syndrome extraction when $n=49$.

\section{Conclusion\label{sec:conclusion}}

I have described a general algorithm for generating thresholds provided that ancillae with independent, identically distributed errors are available as a resource.  My approach applies to most fault-tolerant procedures employing CSS codes.  It relies on the fact that nearly all elements of such a procedure are homogeneous, that is, transversal with identical components.  Inhomogeneous elements are either eliminated, as for classical syndrome processing and the application of recovery unitaries, or, in the case of ancillae, replaced with homogeneous equivalents.  This allows me to calculate the probability of failure for encoded gates in terms of the error probabilities associated with a single strand of the encoded blocks.  In the limit that the number of encoding qubits approaches infinity, a criterion for success becomes simply that the probability of finding an error never exceed the fraction of the encoded qubits on which said error can be corrected.  When this is satisfied, it is possible, in the limit of infinite block size, to compute indefinitely, and our base error rates are, by definition, below threshold.

The value of considering thresholds for homogeneous ancillae is that they can easily be calculated for a variety of fault-tolerant procedures and error models, thereby providing a relatively simple metric for comparison.  Section~\ref{sec:examples} includes thresholds for computation for three fault-tolerant procedures and four error models.  One of the procedures considered is based on a method of telecorrection used by Knill, while the other two are variations, in that the number of syndrome extractions is fixed, on Steane's approach to achieving fault tolerance.  The error models considered are a full depolarizing error model, a depolarizing error model with increased measurement errors, a depolarizing error model for two-qubit gates exclusively, and a restricted two-qubit-gate error model.  Holding the total probability of an error constant, small improvements are observed in the threshold for certain choices of the two-qubit-gate error model.  For the procedures examined, the threshold increases by less than a factor of two, but larger gains may be achievable using fault-tolerant procedures tailored to a specific error model.  With regard to comparisons between procedures, the single-coupling Steane procedure is shown to outperform the double-coupling procedure when two-qubit depolarizing errors dominate, but the double-coupling Steane procedure does substantially better when measurement errors are likely.  I also find that Knill's approach outperforms that of Steane for all error models considered, a conclusion that is likely to hold so long as correlated ancillary errors are rare and the ancillae needed for Knill's method are not appreciably more error prone than those employed by Steane.

Idealized thresholds aside, my algorithm is useful as a means of approximately computing the logical error rate for a single level of encoding, which is an established method of estimating the threshold for quantum computation.  The two treatments yield similar outcomes because numerical estimates of the encoded error rate typically prepare ancillae in a way that maximizes their quality at the cost of additional resource overhead.  Ancillae prepared in this manner have error distributions approximating my ideal of independent, identically distributed errors.  The basic algorithm uses the infinite limit to obtain simple analytic results, but an alternative (and less rigorous) algorithm for finite codes is described in \S\ref{sec:finiteCodes}.  Both methods were shown to yield results in rough accordance with the depolarizing threshold determined by Reichardt~\cite{Reichardt04} for the $[[49,1,9]]$ code.  For telecorrection, in the limit $n\rightarrow\infty$, my estimate of the depolarizing threshold was consistent with the range of values determined by Knill~\cite{Knill04}.

I provide Table~\ref{tab:encodedErrorRates} for researchers who wish to estimate the threshold of the Steane or Knill method given a particular error model.  For those interested in analyzing different fault-tolerant methods, my Mathematica program is available at \href{http://info.phys.unm.edu/~beastin}{http://info.phys.unm.edu/\~{}beastin}\ .

Much further work remains to be done on this subject.  One topic of interest is the tailoring of fault-tolerant procedures to the error model.  It was my initial hope that restricted error models would greatly improve the tolerable error rate.  While this was not observed in existing fault-tolerant procedures, the possibility remains that tailored procedures could improve the threshold.  A second possibility is the extension of my analysis to include memory errors, which promises to be a straightforward, if unbeautiful, endeavor.  The most valuable addition, however, would be to explicitly define methods of ancilla construction and determine the degree to which they differ from my ideal.  Constructing ancillae to my specifications is an extremely difficult problem, but one whose solution would have a strong impact on the theory of quantum computing in general and this work in particular.  A scalable method for producing ancillae with independent, identically distributed errors would enable the algorithm presented here to be employed for the calculation of rigorous lower bounds on the threshold without any caveats about idealized resources.

\begin{acknowledgments}
Suggestions and criticisms regarding this work were provided by Andrew
Silberfarb, Steven Flammia, Andrew Landahl, and Jim Harrington.  John
Preskill graciously pointed out the perils of ancilla construction.
Carlton Caves and Ivan Deutsch supplied inspiration and guidance in all
stages of the project.  I gratefully acknowledge these contributions as
well as funding received from ARO Contract No.~W911NF-04-1-0242 and an
associated QuaCGR Fellowship.
\end{acknowledgments}

%\bibliography{../../citations}

\begin{thebibliography}{23}
\expandafter\ifx\csname natexlab\endcsname\relax\def\natexlab#1{#1}\fi
\expandafter\ifx\csname bibnamefont\endcsname\relax
  \def\bibnamefont#1{#1}\fi
\expandafter\ifx\csname bibfnamefont\endcsname\relax
  \def\bibfnamefont#1{#1}\fi
\expandafter\ifx\csname citenamefont\endcsname\relax
  \def\citenamefont#1{#1}\fi
\expandafter\ifx\csname url\endcsname\relax
  \def\url#1{\texttt{#1}}\fi
\expandafter\ifx\csname urlprefix\endcsname\relax\def\urlprefix{URL }\fi
\providecommand{\bibinfo}[2]{#2}
\providecommand{\eprint}[2][]{\url{#2}}

\bibitem[{\citenamefont{Reichardt}(2004)}]{Reichardt04}
\bibinfo{author}{\bibfnamefont{B.~W.} \bibnamefont{Reichardt}},
  \emph{\bibinfo{title}{Improved ancilla preparation scheme increases
  fault-tolerant threshold}} (\bibinfo{year}{2004}), \eprint{quant-ph/0406025}.

\bibitem[{\citenamefont{Knill}(2005{\natexlab{a}})}]{Knill04}
\bibinfo{author}{\bibfnamefont{E.}~\bibnamefont{Knill}},
  \bibinfo{journal}{Nature} \textbf{\bibinfo{volume}{434}}, \bibinfo{pages}{39}
  (\bibinfo{year}{2005}{\natexlab{a}}), \eprint{quant-ph/0410199}.

\bibitem[{\citenamefont{Knill}(2005{\natexlab{b}})}]{Knill05}
\bibinfo{author}{\bibfnamefont{E.}~\bibnamefont{Knill}},
  \bibinfo{journal}{Phys. Rev. A} \textbf{\bibinfo{volume}{71}},
  \bibinfo{eid}{042322} (pages~\bibinfo{numpages}{7})
  (\bibinfo{year}{2005}{\natexlab{b}}), \eprint{quant-ph/0312190}.

\bibitem[{\citenamefont{Nielsen and Chuang}(2000)}]{NielsenChuang}
\bibinfo{author}{\bibfnamefont{M.~A.} \bibnamefont{Nielsen}} \bibnamefont{and}
  \bibinfo{author}{\bibfnamefont{I.~L.} \bibnamefont{Chuang}},
  \emph{\bibinfo{title}{Quantum computation and quantum information}}
  (\bibinfo{publisher}{Cambridge University Press},
  \bibinfo{address}{Cambridge, England}, \bibinfo{year}{2000}).

\bibitem[{\citenamefont{Preskill}()}]{PreskillNotes}
\bibinfo{author}{\bibfnamefont{J.}~\bibnamefont{Preskill}},
  \emph{\bibinfo{title}{Lecture notes for physics 229: Quantum information and
  computation}}, \urlprefix\url{http://www.iqi.caltech.edu}.

\bibitem[{\citenamefont{Buhrman et~al.}(2006)\citenamefont{Buhrman, Cleve,
  Laurent, Linden, Schrijver, and Unger}}]{Buhrman06}
\bibinfo{author}{\bibfnamefont{H.}~\bibnamefont{Buhrman}},
  \bibinfo{author}{\bibfnamefont{R.}~\bibnamefont{Cleve}},
  \bibinfo{author}{\bibfnamefont{M.}~\bibnamefont{Laurent}},
  \bibinfo{author}{\bibfnamefont{N.}~\bibnamefont{Linden}},
  \bibinfo{author}{\bibfnamefont{A.}~\bibnamefont{Schrijver}},
  \bibnamefont{and} \bibinfo{author}{\bibfnamefont{F.}~\bibnamefont{Unger}},
  \emph{\bibinfo{title}{New limits on fault-tolerant quantum computation}}
  (\bibinfo{year}{2006}), \eprint{quant-ph/0604141}.

\bibitem[{\citenamefont{Razborov}(2003)}]{Razborov03}
\bibinfo{author}{\bibfnamefont{A.~A.} \bibnamefont{Razborov}},
  \bibinfo{journal}{Quantum Information and Computation}
  \textbf{\bibinfo{volume}{4}}, \bibinfo{pages}{222} (\bibinfo{year}{2003}),
  \eprint{quant-ph/0310136}.

\bibitem[{\citenamefont{Aliferis et~al.}(2006)\citenamefont{Aliferis,
  Gottesman, and Preskill}}]{Aliferis06}
\bibinfo{author}{\bibfnamefont{P.}~\bibnamefont{Aliferis}},
  \bibinfo{author}{\bibfnamefont{D.}~\bibnamefont{Gottesman}},
  \bibnamefont{and} \bibinfo{author}{\bibfnamefont{J.}~\bibnamefont{Preskill}},
  \bibinfo{journal}{Quantum Information and Computation}
  \textbf{\bibinfo{volume}{6}}, \bibinfo{pages}{97} (\bibinfo{year}{2006}),
  \eprint{quant-ph/0504218}.

\bibitem[{\citenamefont{Reichardt}(2005)}]{Reichardt05}
\bibinfo{author}{\bibfnamefont{B.~W.} \bibnamefont{Reichardt}},
  \emph{\bibinfo{title}{Fault-tolerance threshold for a distance-three quantum
  code}} (\bibinfo{year}{2005}), \eprint{quant-ph/0509203}.

\bibitem[{\citenamefont{Svore et~al.}(2005)\citenamefont{Svore, Terhal, and
  DiVincenzo}}]{Svore05}
\bibinfo{author}{\bibfnamefont{K.~M.} \bibnamefont{Svore}},
  \bibinfo{author}{\bibfnamefont{B.~M.} \bibnamefont{Terhal}},
  \bibnamefont{and} \bibinfo{author}{\bibfnamefont{D.~P.}
  \bibnamefont{DiVincenzo}}, \bibinfo{journal}{Phys. Rev. A}
  \textbf{\bibinfo{volume}{72}}, \bibinfo{eid}{022317}
  (pages~\bibinfo{numpages}{17}) (\bibinfo{year}{2005}),
  \eprint{quant-ph/0410047}.

\bibitem[{\citenamefont{Svore et~al.}(2006)\citenamefont{Svore, Cross, Chuang,
  and Aho}}]{Svore06}
\bibinfo{author}{\bibfnamefont{K.~M.} \bibnamefont{Svore}},
  \bibinfo{author}{\bibfnamefont{A.~W.} \bibnamefont{Cross}},
  \bibinfo{author}{\bibfnamefont{I.~L.} \bibnamefont{Chuang}},
  \bibnamefont{and} \bibinfo{author}{\bibfnamefont{A.~V.} \bibnamefont{Aho}},
  \bibinfo{journal}{Quantum Information and Computation}
  \textbf{\bibinfo{volume}{6}}, \bibinfo{pages}{193} (\bibinfo{year}{2006}),
  \eprint{quant-ph/0508176}.

\bibitem[{\citenamefont{Eastin}()}]{EastinResidualError}
\bibinfo{author}{\bibfnamefont{B.}~\bibnamefont{Eastin}}, \bibinfo{note}{in
  preparation}.

\bibitem[{\citenamefont{Steane}(1996)}]{Steane96b}
\bibinfo{author}{\bibfnamefont{A.}~\bibnamefont{Steane}},
  \bibinfo{journal}{Proc. R. Soc. A} \textbf{\bibinfo{volume}{452}},
  \bibinfo{pages}{2551} (\bibinfo{year}{1996}), \eprint{quant-ph/9601029}.

\bibitem[{\citenamefont{Calderbank and Shor}(1996)}]{Calderbank96}
\bibinfo{author}{\bibfnamefont{A.~R.} \bibnamefont{Calderbank}}
  \bibnamefont{and} \bibinfo{author}{\bibfnamefont{P.~W.} \bibnamefont{Shor}},
  \bibinfo{journal}{Phys. Rev. A} \textbf{\bibinfo{volume}{54}},
  \bibinfo{pages}{1098} (\bibinfo{year}{1996}), \eprint{quant-ph/9512032}.

\bibitem[{\citenamefont{Gottesman and Preskill}(2001)}]{Gottesman01}
\bibinfo{author}{\bibfnamefont{D.}~\bibnamefont{Gottesman}} \bibnamefont{and}
  \bibinfo{author}{\bibfnamefont{J.}~\bibnamefont{Preskill}},
  \bibinfo{journal}{Phys. Rev. A} \textbf{\bibinfo{volume}{63}},
  \bibinfo{pages}{022309} (\bibinfo{year}{2001}), \eprint{quant-ph/0008046}.

\bibitem[{\citenamefont{Hamada}(2004)}]{Hamada04}
\bibinfo{author}{\bibfnamefont{M.}~\bibnamefont{Hamada}}, \bibinfo{journal}{J.
  Phys. A} \textbf{\bibinfo{volume}{37}}, \bibinfo{pages}{8303}
  (\bibinfo{year}{2004}), \eprint{quant-ph/0308029}.

\bibitem[{\citenamefont{Aharonov and Ben-Or}(1999)}]{Aharonov99}
\bibinfo{author}{\bibfnamefont{D.}~\bibnamefont{Aharonov}} \bibnamefont{and}
  \bibinfo{author}{\bibfnamefont{M.}~\bibnamefont{Ben-Or}},
  \emph{\bibinfo{title}{Fault-tolerant quantum computation with constant error
  rate}} (\bibinfo{year}{1999}), \eprint{quant-ph/9906129}.

\bibitem[{\citenamefont{Emanuel~Knill}(1998)}]{Knill98}
\bibinfo{author}{\bibfnamefont{W.~Z.} \bibnamefont{Emanuel~Knill},
  \bibfnamefont{Raymond~Laflamme}}, \bibinfo{journal}{Proc. R. Soc. A}
  \textbf{\bibinfo{volume}{454}}, \bibinfo{pages}{365} (\bibinfo{year}{1998}),
  \eprint{quant-ph/9702058}.

\bibitem[{\citenamefont{Preskill}(1998)}]{Preskill97}
\bibinfo{author}{\bibfnamefont{J.}~\bibnamefont{Preskill}},
  \emph{\bibinfo{title}{Fault-tolerant Quantum Computation}}
  (\bibinfo{publisher}{World Scientific}, \bibinfo{year}{1998}),
  \eprint{quant-ph/9712048}.

\bibitem[{\citenamefont{Steane}(1998)}]{Steane98}
\bibinfo{author}{\bibfnamefont{A.~M.} \bibnamefont{Steane}},
  \bibinfo{journal}{Fortsch. Phys.} \textbf{\bibinfo{volume}{46}},
  \bibinfo{pages}{443} (\bibinfo{year}{1998}), \eprint{quant-ph/9708021}.

\bibitem[{\citenamefont{Steane}(2003)}]{Steane03}
\bibinfo{author}{\bibfnamefont{A.~M.} \bibnamefont{Steane}},
  \bibinfo{journal}{Physical Review A (Atomic, Molecular, and Optical Physics)}
  \textbf{\bibinfo{volume}{68}}, \bibinfo{eid}{042322}
  (pages~\bibinfo{numpages}{19}) (\bibinfo{year}{2003}),
  \eprint{quant-ph/0207119}.

\bibitem[{\citenamefont{Zalka}(1997)}]{Zalka96}
\bibinfo{author}{\bibfnamefont{C.}~\bibnamefont{Zalka}},
  \emph{\bibinfo{title}{Threshold estimate for fault tolerant quantum
  computation}} (\bibinfo{year}{1997}), \eprint{quant-ph/9612028}.

\bibitem[{\citenamefont{DiVincenzo et~al.}(1998)\citenamefont{DiVincenzo, Shor,
  and Smolin}}]{DiVincenzo98}
\bibinfo{author}{\bibfnamefont{D.~P.} \bibnamefont{DiVincenzo}},
  \bibinfo{author}{\bibfnamefont{P.~W.} \bibnamefont{Shor}}, \bibnamefont{and}
  \bibinfo{author}{\bibfnamefont{J.~A.} \bibnamefont{Smolin}},
  \bibinfo{journal}{Phys. Rev. A} \textbf{\bibinfo{volume}{57}},
  \bibinfo{pages}{830} (\bibinfo{year}{1998}), \eprint{quant-ph/9706061}.

\end{thebibliography}

\appendix

\section{\texorpdfstring{$\pi/4$ Rotation}{T Gate}\label{app:pi4gate}}

The $\pi/4$ rotation, $T$, is not a member of the Clifford group.  This
means that $T$ does not take Pauli strings to Pauli strings under
conjugation.  A logical $T$ gate would also have this property, thereby
confounding the simple error propagation routine used in this paper.  For
CSS codes, however, the encoded $T$ gate is not applied directly; instead,
the logical state
\begin{equation}
\ket{\bar{\Theta}}=\bar{T}\ket{\bar{+}}=\frac{1}{\sqrt{2}}\left(\ket{\bar{0}}+e^{i\pi/4}\ket{\bar{1}}\right)
\end{equation}
is prepared and the data qubit is effectively teleported under the $\bar{T}$ gate using the following encoded circuit:
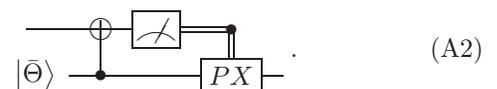
\begin{equation}
\begin{array}{c}
\Qcircuit @R=.4em @C=.8em {
& \qw & \qw & \targ & \meter & \control \cw \cwx[1] \\
& & \lstick{\ket{\bar{\Theta}}} & \ctrl{-1} & \qw & \gate{PX} & \qw
}
\end{array}
.
\end{equation}
If necessary, $P$ can be implemented in the same way as $T$; otherwise the gates applied are Clifford operations (both in terms of the encoding and its constituent qubits), so error propagation proceeds without a hitch.

There is a modest sleight of hand here in that we would not, if we so desired, be able to apply error propagation during the construction of $\ket{\bar{\Theta}}$.  For my purposes this is unimportant since I do not model the construction of ancillae.

\end{document}